 \documentclass[preprint,aps]{revtex4}
 \newcommand{\SO}{\mathrm{SO}}
 \newcommand{\SU}{\mathrm{SU}}
 \newcommand{\unit}{\mathrm{U}}
\usepackage{graphicx}
 
\begin{document}
 \newcommand{\Qed}{\rule{2.5mm}{3mm}}
 \newcommand{\balpha}{\mbox{\boldmath {$\alpha$}}}
 \draft 
%
%
\title{On the origin of families of fermions and their mass matrices}
\author{ A. Bor\v stnik Bra\v ci\v c\\
 Educational Faculty, University of Ljubljana,
 Kardeljeva plo\v s\v cad 17, 1000 Ljubljana\
and\\
N.S. Manko\v c Bor\v stnik\\
Department of Physics, University of
Ljubljana, Jadranska 19, 1000 Ljubljana\\
}
\date{\today}

\begin{abstract} 
We are proposing a new way of describing families of quarks and leptons, using the approach 
unifying all the internal degrees of freedom, proposed by one of 
us\cite{norma92,norma93,normasuper94,norma95,norma97,pikanormaproceedings1,holgernorma00,norma01,%
pikanormaproceedings2,Portoroz03}. Spinors, living in $d \;(=1+13)-$dimensional
space, carry in this approach only the spin and interact with only the gravity through 
vielbeins and two kinds of the spin connection fields - the gauge fields of the Poincar\' e  
group ($p^a, S^{ab}$) and the second kind of the Clifford algebra objects ($\tilde{S}^{ab}$). 
All the quarks and the leptons of one family appear in one Weyl representation of a chosen 
handedness of the Lorentz group, if analyzed with respect to the Standard model gauge 
groups: the right handed (with respect to $SO(1,3)$) weak chargeless 
quarks and leptons and the left handed weak charged quark and leptons. 
A part of the starting Lagrange density of a Weyl spinor in $d=1+13$ 
transforms right handed quarks and leptons 
into left handed quarks and leptons manifesting as the Yukawa 
couplings of the Standard model. The second kind of Clifford algebra objects  generates 
families of quarks and leptons and contributes to diagonal and off diagonal Yukawa couplings. 
The approach predicts an even number of families, treating leptons and quarks equivalently. 
In this paper we investigate within this approach 
the appearance of the Yukawa couplings within one family of quarks and leptons as well 
as among the families (without assuming any Higgs fields). We present the mass matrices 
for four families and investigate whether our way of generating families might explain the origin 
of families of quarks and leptons as well as their observed properties - the masses and the 
mixing matrices. Numerical results are presented in the paper following this one\cite{matjazdragannorma}.
\end{abstract}

\maketitle

\section{Introduction}
\label{introduction}

The Standard model of the electroweak and strong interactions (extended by the inclusion of the massive 
neutrinos) fits well all the existing experimental data. It assumes around 25 parameters and constraints, 
the origin of which is not yet understood. Questions like: Why has Nature chosen 
$SU(3)\times SU(2)\times U(1)$ to describe the charges of spinors and $SO(1,3)$ to describe 
the spin of spinors?, 
Why are the left handed spinors  weak charged, while the right handed spinors are weak chargeless?,
Where do the Yukawa couplings (together with the weak scale and the families of quarks and leptons)
come from?, and many others, remain unanswered.

The advantage of the approach, unifying spins and 
charges\cite{norma92,norma93,normasuper94,norma95,norma97,%
pikanormaproceedings1,holgernorma00,norma01,pikanormaproceedings2,Portoroz03}, is that 
it might offer  possible 
answers to the open questions of the Standard electroweak model. We demonstrated in 
references\cite{pikanormaproceedings1,%
norma01,pikanormaproceedings2,Portoroz03} that a left handed $SO(1,13)$ 
Weyl spinor multiplet includes, if the representation is interpreted
in terms of the subgroups $SO(1,3)$, $SU(2)$, $SU(3)$ and the sum of the two $U(1)$'s,  
all the spinors of
the Standard model - that is the left handed $SU(2)$ doublets and the right handed  $SU(2)$ 
singlets of (with the group  $SU(3)$ charged) quarks and  (chargeless) leptons.
Right handed neutrinos -  weak and hyper chargeless - are also included.
In the gauge theory of gravity (in our case in $d=(1+13)$-dimensional space), 
the Poincar\' e group is gauged, 
leading to spin connections and vielbeins, which  determine the gravitational 
field\cite{mil,norma93,norma01}.
There are  vielbein fields and spin connection fields, which might manifest - after 
the appropriate compactification 
(or some other kind of making the rest of d-4 space unobservable at 
low energies) - in the four dimensional 
space-time as all the gauge fields of the known charges, as well as the Yukawa couplings within 
each family. No additional Higgs field is needed to generate masses of families and 
to ''dress'' the right handed spinors with the weak charge. It is a part of the 
starting Lagrangean in $d\ge1+13$, which manifests in $d=1+3$ as the Yukawa coupling and does 
what the Higgs does in the Standard model. 
If assuming a second  kind of the Clifford algebra objects,  
the corresponding gauge fields manifest as the Yukawa couplings among families 
(contributing also to 
Yukawa couplings within each family).  

In the refs.\cite{pikanormaproceedings2,Portoroz03,
holgernorma02,technique03,astridragannorma,bmnBled04} it was shown, 
that the approach unifying spins and charges might explain the Yukawa couplings if 
an appropriate break of both symmetries, connected with the two kinds of the 
Clifford algebra objects, appears.  
 An even number of families is predicted, in particular,
the fourth family of quarks and leptons might appear under certain conditions at low energies 
in agreement with\cite{okun}.

The approach seems to have, like all the Kaluza-Klein-like theories, a very 
serious disadvantage, namely that there might not exist  any massless, 
mass protected spinors, which are, after the break of symmetries, 
chirally coupled to the desired (Kaluza-Klein) gauge fields\cite{witten}. 
This would mean that there are no observable spinors at low energies. 
Since the idea that it is only one internal degree of freedom - 
the spin - and the Kaluza-Klein idea that the gravity is the only gauge field,  
are beautiful and attractive,
we have tried hard to find any  example
, which would give hope to Kaluza-Klein-like 
theories by demonstrating that a kind of a break of symmetries  leads to 
massless, mass protected spinors, 
chirally coupled to the Kaluza-Klein gauge fields, observable at low energies.
We discuss in ref.\cite{holgernorma05} such a case - a toy model of a spinor, 
living in $d(=1+5)$-dimensional space, which breaks into a finite disk with 
the boundary, which allows spinors of only one handedness.
Although not yet realistic, the toy model looks promising.

In the present paper we analyze how do families of quarks and leptons, 
and accordingly also the Yukawa couplings, appear within the 
approach unifying spins and charges. 
We comment on the type of 
contributions to the Yukawa couplings and discuss some general properties 
of the mass matrices, which follow from the assumptions of the approach, 
trying to find out whether the approach could show a possible answer to the questions: 
What is the origin of the families of quarks and leptons?, What does determine the Yukawa couplings?,
Why only the left handed quarks and leptons carry the weak charge?.

Since we do not know, which way of breaking the starting symmetries of the approach is the appropriate 
one and since results of the investigation drastically depend on the way of breaking symmetries 
 and might as well depend on non adiabatic processes following the break of symmetries, 
this paper (and also the paper which follows this one and represent some numerical 
investigations)  can only be understood as an attempt to see whether the 
approach unifying spins and 
charges has a chance to explain the origin of families of quarks and leptons and their 
properties and to which extend might it help to understand the appearance of families and the 
Yukawa couplings.

We are not (yet) performing the  calculations of breaking the symmetry  
$SO(1,13)$ to $SO(1,7) \times U(1) \times SU(3)$  within our approach.  (Some very rough 
estimates can be found in ref.\cite{hnrunBled02}.) The break of symmetries 
influences both kinds of gauge fields, although we can not yet tell indeed in which way.   
Therefore,  we can not tell the strength of
the fields which appear in the Yukawa couplings as "the vacuum expectation values" 
and which lead further to 
$SO(1,3) \times U(1) \times SU(3)$. We only can evaluate (after making some assumptions) several
relations among the spin connection fields. Using then these very preliminary relations and 
the known experimental data, 
we can make a  prediction for the number of families at "physical energies"  and discuss 
 properties of quarks and leptons within this approach: their  masses and mixing matrices. 
Accordingly the results can  be taken only  as a first step in analyzing properties of 
 families of quarks and leptons within {\em the approach unifying spins and charges, 
 which might offers a mechanism for generating families and 
 correspondingly the Yukawa couplings.}  
We shall present some numerical results in the paper, following 
this one. 

In Sect.\ref{lagrangesec} of this paper we present the action  for a Weyl spinor in 
$(1+13)$-dimensional space 
within  our approach and suggest a break of the symmetry $SO(1,13)$ 
to $SO(1,7)  \times SO(6)$ and further. We assume that the break of $SO(1,13)$ to 
$SO(1,7) \times U(1) \times SU(3)$
does lead to massless Weyl spinors with the $U(1)\times SU(3)$ charges. 
The main point of this paper is to demonstrate  that while
one Weyl spinor representation of $SO(1,13)$, if analyzed with respect to subgroups 
$SO(1,3)\times SU(2)\times
U(1)\times U(1)\times SU(3)$, contains all the spinors needed in the Standard model 
(the right handed weak chargeless quarks and leptons
and left handed weak charged quarks and leptons), {\it the starting action for a  Weyl spinor, 
which carries
only (two kinds of) the spin}  {\it and no charges and interacts with only the gravitational field, 
includes the Yukawa couplings}, which  transform 
the right handed weak chargeless spinors into left handed weak charged spinors and contribute to the 
mass terms just as it is suggested by the Standard model, without  assuming  the existence 
of a Higgs weak charged doublet. There are, namely, the generators of the Lorentz transformations 
within the group $SO(1,7)$ in our model ($S^{0s},\;s=7,8$, for example), which take care of what 
in the Standard model the Higgs doublet, together with $\gamma^0$, does.

In Subsect.\ref{break} we comment on a possible break of the starting symmetry $SO(1,13)$ the 
internal symmetry of which is connected by $S^{ab}$, while in
Subsect.\ref{so1,13} of Sect.\ref{lagrangesec} we discuss properties of the group $SO(1,13)$ in 
terms of subgroups, which appear in the Standard electroweak model.
In the same section, Subsect.\ref{technique}, we present briefly the 
technique\cite{holgernorma02,technique03}, 
which turns out to be very helpful when discussing   spinor representations, since it allows to
generate as well as present spinor representations and families of spinor representations 
in a very transparent way. In particular, the technique 
helps to point out very clearly how do the Yukawa couplings appear in our approach. 
In Subsect.\ref{techniquefamilies} we comment on the 
appearance of families within our technique\cite{norma93,technique03}. 

In Sect.\ref{Yukawawithin} we discuss in details within our approach the appearance of 
the Yukawa couplings within one family,
while in Sect.\ref{Yukawafamilies} we  discuss the number of families as well as the Yukawa couplings 
among the families.

In Sect.\ref{example} we present, after making several assumptions and simplifications, 
a possible explicit expression for the mass matrices for four families of quarks and leptons 
in terms of the spin connection fields.


\section{ Weyl spinors in $d= (1+13)$ manifesting families of quarks and leptons in $d= (1+3)$} 
\label{lagrangesec}

We start with a  left handed Weyl spinor  in $(1+13)$-dimensional space. A spinor carries no charges, 
only two kinds of  spins and interacts accordingly with only gauge gravitational 
fields 
- with spin connections and vielbeins. We assume two kinds of the Clifford 
algebra objects defining two kinds of the generators of the Lorentz algebra and allow accordingly 
two kinds of gauge fields\cite{norma92,norma93,normasuper94,norma95,norma97,%
pikanormaproceedings1,holgernorma00,norma01,pikanormaproceedings2,Portoroz03}. One kind is the ordinary
gauge field (gauging the Poincar\' e symmetry in $d=1+13$). The corresponding spin connection field
appears for spinors as a gauge field of $S^{ab}= \frac{1}{4} (\gamma^a \gamma^b - \gamma^b \gamma^a)$, 
where $\gamma^a$ are the ordinary Dirac operators. 
The contribution of these fields to the mass matrices manifests in only the diagonal terms  
(connecting right handed weak chargeless quarks or leptons with left handed weak charged partners within 
one family of spinors). 

The second kind of gauge fields is in our approach responsible for the appearance of families and 
consequently for the Yukawa couplings
among families of spinors (contributing also to diagonal matrix elements) and will be used 
in this paper to explain the origin of the families of quarks and leptons. 
The corresponding spin connection fields appear for spinors as a gauge field 
of $\tilde{S}^{ab}$ ($\tilde{S}^{ab} = \frac{1}{2} (\tilde{\gamma}^a \tilde{\gamma}^b-
\tilde{\gamma}^b \tilde{\gamma}^a)$) with $\tilde{\gamma}^a$, which are 
the Clifford algebra objects\cite{norma93,technique03}, like $\gamma^a$, but anticommute with $\gamma^a$.

Accordingly we write the action for a Weyl (massless) spinor  
in $d(=1+13)$ - dimensional space as follows\footnote{Latin indices  
$a,b,..,m,n,..,s,t,..$ denote a tangent space (a flat index),
while Greek indices $\alpha, \beta,..,\mu, \nu,.. \sigma,\tau ..$ denote an Einstein 
index (a curved index). Letters  from the beginning of both the alphabets
indicate a general index ($a,b,c,..$   and $\alpha, \beta, \gamma,.. $ ), 
 from the middle of both the alphabets   
 the observed dimensions $0,1,2,3$ ($m,n,..$ and $\mu,\nu,..$), indices from the bottom of the alphabets
 indicate the compactified dimensions ($s,t,..$ and $\sigma,\tau,..$). We assume the signature $\eta^{ab} =
diag\{1,-1,-1,\cdots,-1\}$.
}
\begin{eqnarray}
S &=& \int \; d^dx \; {\mathcal L}
\nonumber\\
{\mathcal L} &=& \frac{1}{2} (E\bar{\psi}\gamma^a p_{0a} \psi) + h.c. = \frac{1}{2} 
(E\bar{\psi} \gamma^a f^{\alpha}{}_a p_{0\alpha}\psi) + h.c. ,
\nonumber\\
p_{0\alpha} &=& p_{\alpha} - \frac{1}{2}S^{ab} \omega_{ab\alpha} - 
\frac{1}{2}\tilde{S}^{ab} \tilde{\omega}_{ab\alpha}.
\label{lagrange}
\end{eqnarray}
Here $f^{\alpha}{}_a$ are  vielbeins (inverse to the gauge field of the generators of translations  
$e^{a}{}_{\alpha}$, $e^{a}{}_{\alpha} f^{\alpha}{}_{b} = \delta^{a}_{b}$,
$e^{a}{}_{\alpha} f^{\beta}{}_{a} = \delta_{\alpha}{}^{\beta}$),
with $E = det(e^{a}{}_{\alpha})$, while  
$\omega_{ab\alpha}$ and $\tilde{\omega}_{ab\alpha} $ are the two kinds of the spin 
connection fields, the gauge 
fields of $S^{ab}$ and $\tilde{S}^{ab}$, respectively, corresponding to the two kinds of 
the Clifford algebra 
objects\cite{holgernorma02,Portoroz03}, namely $\gamma^a$ and $\tilde{\gamma}^{a}$, with the property 
$\{\gamma^a,\tilde{\gamma}^b\}_+ =0$, which leads to 
$\{ S^{ab}, \tilde{S}^{cd}\}_-=0$.  We shall discuss the properties 
of these two kinds of $\gamma^a$'s in Subsects.\ref{technique} and \ref{techniquefamilies}.

To see that  one Weyl spinor in $d=(1+13)$ with the spin as the only internal degree of freedom, can
manifest  in
four-dimensional ''physical'' space  as the ordinary ($SO(1,3)$) spinor with all the known charges 
of one family of  quarks and leptons of the Standard model, one has to analyze one
Weyl spinor (we make a choice of the left handed one) representation in terms of the subgroups 
$SO(1,3) \times
U(1) \times SU(2) \times SU(3)$. We shall do this in Subsect.\ref{so1,13} of this section. 
(The reader can see this analyses in several references, like the one in\cite{Portoroz03}.)

To see that the Yukawa couplings are the part of the starting Lagrangean of Eq.(\ref{lagrange}), we rewrite
the Lagrangean in  Eq.(\ref{lagrange}) as follows\cite{Portoroz03} 
\begin{eqnarray}
{\mathcal L} &=& \bar{\psi}\gamma^{m} (p_{m}- \sum_{A,i}\; g^{A}\tau^{Ai} A^{Ai}_{m}) \psi 
+ \nonumber\\
& &  \sum_{s=7,8}\; 
\bar{\psi} \gamma^{s} p_{0s} \; \psi + {\rm the \;rest}.
\label{yukawa}
\end{eqnarray}
Index $A$ determines the charge groups ($SU(3), SU(2)$ and the two $U(1)$'s), index $i$ determines
the generators within one charge group. $\tau^{Ai}$ denote the generators of the charge groups 
(expressible\cite{norma01} in terms of $S^{st},\; s,t \in 5,6,..,14$), while $A^{Ai}_{m}, m=0,1,2,3,$ 
denote the corresponding
gauge fields (expressible in terms of $\omega_{st m}$).

The second term can be rewitten in terms of the kinetic part ($\psi^{\dagger} \gamma^0\gamma^s p_s \psi$) 
and the part $ - \psi^{\dagger} \gamma^0\gamma^{s} S^{ t t'} f^{\sigma}_s \omega_{t t' \sigma} \psi -   
\psi^{\dagger} \gamma^0 \gamma^{s} \tilde{S}^{t t'} f^{\sigma}_s \tilde{\omega}_{t t' \sigma} \psi $, which 
looks like a  mass term (also the kinetic term, if nonzero, contributes to the mass term), since 
$f^{\sigma}_s \omega_{t t' \sigma}$ and $f^{\sigma}_s \tilde{\omega}_{t t' \sigma}, $ 
$ s,t \in 5,6,7,8,\; \sigma \in (5),(6),(7),(8))$, behave in $d(=1+3)-$
dimensional  space like scalar fields, while the operator $\gamma^0\gamma^{s}, s=7,8$, for example, 
transforms a right handed
weak chargeless spinor (for example $e_R$) into a left handed weak charged spinor (in this case to $e_L$), 
without changing the spin in $d=1+3$ (Subsect.\ref{technique}, Eq.(\ref{graphgammaaction}) 
and the third and the fifth 
row of Table II or the fourth and the sixth row of the same table) - just
what the Yukawa couplings with the Higgs doublet included do in the Standard model formulation. 
The reader will 
find the detail explanation in Subsects.\ref{technique},\ref{techniquefamilies}. 
It should be pointed out that no Higgs weak charge doublet is needed here, 
as $S^{0s}, s=7,8$ does its job.

One can always rewrite the Lagrangean from Eq.(\ref{lagrange}) in the way of Eq.(\ref{yukawa}).
The question is, of course, what are the terms, which are in Eq.(\ref{yukawa}) written
under  ''the rest'' and whether they can be assumed as negligible at ''low energy world''.   
We have no proof that any break of symmetry, presented in Subsect.\ref{break},
leads to such an effective Lagrangean, which would after the first break (or several successive breaks) of
the starting symmetry of $SO(1,13)$ manifest any massless spinors, which would then, after further breaks, 
manifest in the "physical space" the masses corresponding to  
the Yukawa couplings  of Eq.(\ref{yukawa}), while all the rest terms are negligibly small. 
We just assume instead, that we start with the Lagrangean of
Eq.(\ref{yukawa}) and then study properties of the system, described by such a Lagrange density  
at ''physical'' energies. 

We also would like to point out that  the fact that the generators of families of spinors 
$\tilde{S}^{ab}$ and the generators of the Lorentz transformations of spinors $S^{ab}$ commute 
($\{\tilde{S}^{ab},S^{ab}\}_- =0 $), suggests  that most of properties of 
quarks and leptons must be the same within this approach. There are namely only the generators 
of families, which define off diagonal elements of the Yukawa couplings. But they do not at all  
distinguish among quarks and leptons. Since also in the diagonal matrix elements differ quarks 
and leptons in only one parameter times the identity, the question arises: 
What is then the reason for so different mixing matrices of 
quarks and leptons as observed? Might it be that there are the nonperturbative effects 
(like in the hadron case when quarks ''dress nonadiabatically'' 
into the clouds of quarks and antiquarks and the gluon field before forming a hadron) which are 
responsible 
for  so different properties of quarks and leptons? Could instead be that very peculiar breaks  
of symmetries cause the difference in off diagonal matrix elements for quarks and leptons? Or 
one must take the appearance of the Majorana fermions into account?  
The approach by itself gives different off diagonal 
elements of mass matrices for $u$-quarks and $d$-quarks, and for $\nu$ and electrons (although 
it still relates them). 
We shall discuss this point later in this paper, as well as in the paper following   this one.


\subsection{Break of symmetries}
\label{break}

There are several ways of breaking  the group $SO(1,13)$ down to subgroups of the Standard model.
(One of) the most probable breaks, suggested by the approach unifying spins and charges, is the  
following one 
\[
 \begin{array}{c}
 \begin{array}{c}
 \underbrace{%
 \begin{array}{rrcll}
  & & \SO(1,13) \\
  & & \downarrow \\
  & & \SO(1,7) \otimes \SU(3) \otimes U(1) \\
 & & & \\
 \end{array}} \\
 \end{array}\\
 \downarrow \\
 \SO(1,3)\otimes\unit(1)\otimes\SU(3)\\
 \end{array}
 \]
We start from a massless left handed Weyl spinor in $d=1+13$. We assume that the first break of symmetries 
leads again to massless spinors in $d=1+7$, chirally coupled with the $SU(3)$ and $U(1)$ charge 
to the corresponding fields, which follow from the spin connection and vielbein fields in $d=1+13$. 
(The reader can find more about this kind of breaking the 
starting symmetry in ref.\cite{hnrunBled02}.) We have no justification for such an assumption 
(except that we have shown on  one toy model\cite{holgernorma05} that in that very special case such an 
assumption is justified). And we have no calculation, which would help to guess the strength of 
the ''vacuum expectation values'' of the fields.  
The Yukawa like terms themselves then break further the symmetry, ending up with the 
''physical'' degrees of freedom. An additional non yet solved problem is, how does the break of 
symmetries influences 
the part $\tilde{S}^{st}\tilde{\omega}_{sts^,}$.


\subsection{Spin and charges of  one left handed Weyl representation of SO(1,13)}
\label{so1,13}

We discuss in this subsection the properties of one Weyl spinor representation when 
analysing the representation in terms of subgroups of the group $SO(1,13).$

The group $SO(1,13)$ of the rank $7$ has as possible subgroups the groups 
$SO(1,3)$ (the  ''complexified''
$SU(2) \times  SU(2)$),  $SU(2), SU(3)$ and the two $U(1)$'s, with the sum of the ranks of all 
these subgroups  equal to $7$. These subgroups 
 are  candidates for describing the spin, the weak charge, the colour charge and the two hyper charges, 
 respectively (only one is needed in the Standard model). 
 The generators of these groups  can be written in terms of the generators $S^{ab}$ as follows 
\begin{eqnarray}
\tau^{Ai} = \sum_{a,b} \;c^{Ai}{ }_{ab} \; S^{ab},
\nonumber\\
\{\tau^{Ai}, \tau^{Bj}\}_- = i \delta^{AB} f^{Aijk} \tau^{Ak}.
\label{tau}
\end{eqnarray}
We could count the two $SU(2)$ subgroups of the group $SO(1,3)$ in the same way as the rest of subgroups. 
Instead we shall
use $A=1,2,3,4,$ to represent only the subgroups describing charges  and $f^{Aijk}$ to describe 
the corresponding structure 
constants. Coefficients $c^{Ai}{ }_{ab}$, with $a,b \in \{5,6,...,14\}$, have to be determined so that the 
commutation relations of Eq.(\ref{tau}) 
hold\cite{norma97}.

The weak charge  ($SU(2)$ with the generators $\tau^{1i}$) and  one $ U(1)$ charge (with the generator
$\tau^{21}$)  content of the compact group $SO(4)$
 (a subgroup of $SO(1,13)$) can  be demonstrated when expressing
\begin{eqnarray}
\tau^{11}: = \frac{1}{2} ( {\mathcal S}^{58} - {\mathcal S}^{67} ),\quad
\tau^{12}: = \frac{1}{2} ( {\mathcal S}^{57} + {\mathcal S}^{68} ),\quad
\tau^{13}: = \frac{1}{2} ( {\mathcal S}^{56} - {\mathcal S}^{78} ),
\nonumber\\
\tau^{21}: = \frac{1}{2} ( {\mathcal S}^{56} + {\mathcal S}^{78} ).
\label{su12w}
\end{eqnarray}

To see the colour charge  and one additional $U(1)$ content in the group $SO(1,13)$ we write $\tau^{3i}$ and 
$\tau^{41}$, respectively, in terms of  the generators ${\mathcal S}^{ab}$ 
\begin{eqnarray}
\tau^{31}: &=& \frac{1}{2} ( {\mathcal S}^{9\;12} - {\mathcal S}^{10\;11} ),\quad
\tau^{32}: = \frac{1}{2} ( {\mathcal S}^{9\;11} + {\mathcal S}^{10\;12} ),\quad
\tau^{33}: = \frac{1}{2} ( {\mathcal S}^{9\;10} - {\mathcal S}^{11\;12} ),\quad
\nonumber
\\
\tau^{34}:&=& \frac{1}{2} ( {\mathcal S}^{9\;14} - {\mathcal S}^{10\;13} ),\quad
\tau^{35}: = \frac{1}{2} ( {\mathcal S}^{9\;13} + {\mathcal S}^{10\;14} ),\quad
\tau^{36}: = \frac{1}{2} ( {\mathcal S}^{11\;14} - {\mathcal S}^{12\;13}),\quad 
\nonumber\\
\tau^{37}: &=& \frac{1}{2} ( {\mathcal S}^{11\;13} + {\mathcal S}^{12\;14} ),\quad
\tau^{38}: = \frac{1}{2\sqrt{3}} ( {\mathcal S}^{9\;10} + {\mathcal
S}^{11\;12} - 2{\mathcal S}^{13\;14}),
\nonumber\\
\tau^{41}: &=& -\frac{1}{3}( {\mathcal S}^{9\;10} + {\mathcal S}^{11\;12}
+ {\mathcal S}^{13\;14} ).
\label{su3u1so6}
\end{eqnarray}
To reproduce the Standard model groups one must introduce the two superpositions of the two $U(1)$'s 
generators as follows
\begin{eqnarray}
Y = \tau^{41} + \tau^{21}, \quad  Y' = \tau^{41} - \tau^{21}. 
\label{yyprime}
\end{eqnarray}
The above choice of subgroups of the group $SO(1,13)$ manifests the Standard model charge structure 
of one Weyl spinor of the 
group $SO(1,13)$, with one additional hyper charge. 

We may very similarly proceed also with the generators $\tilde{S}^{ab}$ by assuming that 
a kind of a break makes the starting $SO(1,13)$ group to manifest in terms of some 
$\tilde{\tau}^{\tilde{A}i}$ like 
\begin{eqnarray}
\tilde{\tau}^{\tilde{A}i} = \sum_{a,b} \;\tilde{c}^{\tilde{A}i}{ }_{ab} \; \tilde{S}^{ab},
\nonumber\\
\{\tilde{\tau}^{\tilde{A}i}, \tilde{\tau}^{\tilde{B}j}\}_- = i \delta^{\tilde{A}\tilde{B}} 
\tilde{f}^{\tilde{A}ijk} \tilde{\tau}^{\tilde{A}k}.
\label{tildetau}
\end{eqnarray}
We shall try to guess the way of breaking through the comparison of the results with the 
experimental data in the paper following this one.


\subsection{Spinor representation in terms of Clifford algebra objects}
\label{technique}

In this subsection we  briefly present our technique\cite{holgernorma02} for generating spinor representations
in any dimensional space. The advantage of this technique is simplicity in using it and transparency 
in understanding  detailed properties of spinor representations. We also show how families of spinors enter 
into our approach\cite{norma93,technique03}.

We start by defining two kinds of the Clifford algebra objects, $\gamma^a$ and $\tilde{\gamma^a}$, 
with the properties
\begin{eqnarray}
\{\gamma^a,\gamma^b\}_{+} = 2\eta^{ab} =  \{\tilde{\gamma}^a,\tilde{\gamma}^b\}_{+},
\quad \{\gamma^a,\tilde{\gamma}^b\}_{+} = 0.
\label{clifford}
\end{eqnarray}
The operators $\tilde{\gamma}^a$  are introduced formally as operating on any Clifford algebra object $B$ 
from the left hand side, but they also can be expressed in terms of the  ordinary $\gamma^a$ as  
operating from the right hand side as follows
\begin{eqnarray}
\tilde{\gamma}^a B : = i(-)^{n_B} B \gamma^a,
\label{tildegclifford}
\end{eqnarray}
with $(-)^{n_B} =  +1$ or $-1$, when the object $B$ has a Clifford even or odd character, respectively.

Accordingly two kinds of generators of the Lorentz transformations follow, namely  
$S^{ab}: = (i/4) (\gamma^a \gamma^b - \gamma^b \gamma^a)$ and 
$\tilde{S}^{ab}: = (i/4) (\tilde{\gamma}^a \tilde{\gamma}^b - \tilde{\gamma}^b \tilde{\gamma}^a)$, with the property
$\{ S^{ab},\tilde{S}^{cd}\}_{-}=0$.

We define a basis of spinor representations as  
eigen states of the chosen Cartan subalgebra of the Lorentz algebra $SO(1,13)$,  with  the operators 
$S^{ab}$ and $\tilde{S}^{ab}$ in the two Cartan subalgebra sets, with the same indices in both cases.

By introducing the notation
\begin{eqnarray}
\stackrel{ab}{(\pm i)}: &=& \frac{1}{2}(\gamma^a \mp  \gamma^b),  \quad 
\stackrel{ab}{[\pm i]}: = \frac{1}{2}(1 \pm \gamma^a \gamma^b), \;{\rm for} \; \eta^{aa} \eta^{bb} =-1, \nonumber\\
\stackrel{ab}{(\pm )}: &= &\frac{1}{2}(\gamma^a \pm i \gamma^b),  \quad 
\stackrel{ab}{[\pm ]}: = \frac{1}{2}(1 \pm i\gamma^a \gamma^b), \;{\rm for} \; \eta^{aa} \eta^{bb} =1,
\label{eigensab}
\end{eqnarray}
it can be shown that  
\begin{eqnarray}
S^{ab} \stackrel{ab}{(k)} &=&  \frac{k}{2} \stackrel{ab}{(k)}, \quad 
S^{ab} \stackrel{ab}{[k]}  =  \frac{k}{2} \stackrel{ab}{[k]}, \nonumber\\
\tilde{S}^{ab} \stackrel{ab}{(k)}  &= & \frac{k}{2} \stackrel{ab}{(k)},  \quad 
\tilde{S}^{ab} \stackrel{ab}{[k]}   =   - \frac{k}{2} \stackrel{ab}{[k]}.
\label{eigensabev}
\end{eqnarray}
The above binomials are all ''eigen vectors''  of  the generators $S^{ab}$, as well as of  $\tilde{S}^{ab}$.

We further find 
\begin{eqnarray}
\gamma^a \stackrel{ab}{(k)}&=&\eta^{aa}\stackrel{ab}{[-k]},\quad 
\gamma^b \stackrel{ab}{(k)}= -ik \stackrel{ab}{[-k]}, \nonumber\\
\gamma^a \stackrel{ab}{[k]}&=& \stackrel{ab}{(-k)},\quad \quad \quad
\gamma^b \stackrel{ab}{[k]}= -ik \eta^{aa} \stackrel{ab}{(-k)}
\label{graphgammaaction}
\end{eqnarray}
and
\begin{eqnarray}
\tilde{\gamma^a} \stackrel{ab}{(k)} &=& - i\eta^{aa}\stackrel{ab}{[k]},\quad
\tilde{\gamma^b} \stackrel{ab}{(k)} =  - k \stackrel{ab}{[k]}, \nonumber\\
\tilde{\gamma^a} \stackrel{ab}{[k]} &=&  \;\;i\stackrel{ab}{(k)},\quad \quad \quad
\tilde{\gamma^b} \stackrel{ab}{[k]} =  -k \eta^{aa} \stackrel{ab}{(k)}.
\label{gammatilde}
\end{eqnarray}
Using the following useful relations
\begin{eqnarray}
\stackrel{ab}{(k)}^{\dagger}=\eta^{aa}\stackrel{ab}{(-k)},\quad
\stackrel{ab}{[k]}^{\dagger}= \stackrel{ab}{[k]},
\label{graphher}
\end{eqnarray}
we may define 
\begin{eqnarray}
<\; \stackrel{ab}{(k)}^{\dagger} | \stackrel{ab}{(k)}\;>=1= <\;\stackrel{ab}{[k]}^{\dagger} 
| \stackrel{ab}{[k]}\;>.
\label{scalar}
\end{eqnarray}
We shall later make use  of the relations
\begin{eqnarray}
\stackrel{ab}{(k)}\stackrel{ab}{(k)}& =& 0, \quad \quad \stackrel{ab}{(k)}\stackrel{ab}{(-k)}
= \eta^{aa}  \stackrel{ab}{[k]}, \quad 
\stackrel{ab}{[k]}\stackrel{ab}{[k]} =  \stackrel{ab}{[k]}, \quad \quad
\stackrel{ab}{[k]}\stackrel{ab}{[-k]}= 0, 
 \nonumber\\
\stackrel{ab}{(k)}\stackrel{ab}{[k]}& =& 0,\quad \quad \quad \stackrel{ab}{[k]}\stackrel{ab}{(k)}
=  \stackrel{ab}{(k)}, \quad \quad 
\stackrel{ab}{(k)}\stackrel{ab}{[-k]} =  \stackrel{ab}{(k)},
\quad \quad \stackrel{ab}{[k]}\stackrel{ab}{(-k)} =0,  \quad \quad 
\label{graphbinoms}
\end{eqnarray}
as well as the relations, following from Eqs.(\ref{gammatilde},\ref{graphbinoms}), 
\begin{eqnarray}
\stackrel{ab}{\tilde{(k)}} \stackrel{ab}{(k)}& =& 0, 
\quad \quad \stackrel{ab}{\tilde{(-k)}} \stackrel{ab}{(k)}
= -i \eta^{aa}  \stackrel{ab}{[k]}, 
\quad \stackrel{ab}{\tilde{(-k)}}\stackrel{ab}{[-k]}= i \stackrel{ab}{(-k)},\quad
\stackrel{ab}{\tilde{(k)}} \stackrel{ab}{[-k]} = 0, \nonumber\\
\stackrel{ab}{\tilde{(k)}} \stackrel{ab}{[k]}& =& i \stackrel{ab}{(k)}, \;\;
\stackrel{ab}{\tilde{(-k)}}\stackrel{ab}{[+k]}= 0, \;\;\quad \quad  \quad \stackrel{ab}{\tilde{(-k)}}\stackrel{ab}{(-k)}=0,
 \;\;\stackrel{ab}{\tilde{(k)}}\stackrel{ab}{(-k)} = -i \eta^{aa} \stackrel{ab}{[-k]}.
\label{graphbinomsfamilies}
\end{eqnarray}
Here 
\begin{eqnarray}
\stackrel{ab}{\tilde{(\pm i)}} = \frac{1}{2} (\tilde{\gamma}^a \mp \tilde{\gamma}^b), \quad
\stackrel{ab}{\tilde{(\pm 1)}} = \frac{1}{2} (\tilde{\gamma}^a \pm i\tilde{\gamma}^b), \nonumber\\
\stackrel{ab}{\tilde{[\pm i]}} = \frac{1}{2} (1 \pm \tilde{\gamma}^a \tilde{\gamma}^b), \quad
\stackrel{ab}{\tilde{[\pm 1]}} = \frac{1}{2} (1 \pm i \tilde{\gamma}^a \tilde{\gamma}^b). 
\label{deftildefun}
\end{eqnarray}

The reader should notice that $\gamma^a$'s transform the binomial  
$\stackrel{ab}{(k)}$ into the binomial $\stackrel{ab}{[-k]}$,
whose ''eigen value'' with respect to $S^{ab}$ changes sign, while
$\tilde{\gamma}^a$'s transform the binomial $\stackrel{ab}{(k)}$ 
into $\stackrel{ab}{[k]}$ with unchanged ''eigen value''
with respect to $S^{ab}$. 
We define the operators of  handedness of the group $SO(1,13)$ and of the subgroups $SO(1,3), SO(1,7), 
SO(6)$ and $SO(4)$ as follows
\begin{eqnarray}
\Gamma^{(1,13)} &=&  i 2^{7} \; S^{03} S^{12} S^{56} \cdots S^{13 \; 14}, \quad 
\Gamma^{(1,3)}\;=  - i 2^2 S^{03} S^{12}, \nonumber \\ 
\Gamma^{(1,7)}\;&=&  - i2^{4}  S^{03} S^{12} S^{56} S^{78},\quad \quad \quad
\Gamma^{(1,9)}\; = i2^{5}  S^{03} S^{12} S^{9\;10} S^{11\;12} S^{13 \; 14},\nonumber\\
\Gamma^{(6)}\;\;&=& - 2^3 S^{9 \;10} S^{11\;12} S^{13 \; 14},\quad\quad\;\;\; 
\Gamma^{(4)}\;\;= 2^2 S^{56} S^{78}.
\label{handedness}
\end{eqnarray}

We shall represent one Weyl left handed spinor as products of binomials $\stackrel{ab}{(k)}$ or  
$\stackrel{ab}{[k]}$, which are  ''eigen vectors'' of the members of the Cartan subalgebra set.
We make the following choice of the Cartan subalgebra   set of the algebra $S^{ab}$
\begin{eqnarray}
S^{03}, S^{12}, S^{56}, S^{78}, S^{9 \;10}, S^{11\;12}, S^{13\; 14}.
\label{cartan}
\end{eqnarray}

We are now prepared to make a choice of a starting basic vector  of one Weyl representation of the group 
$SO(1,13)$, which is the
eigen state of all the members of the Cartan subalgebra (Eq.(\ref{cartan})) and is left handed
($\Gamma^{(1,13)} =-1$) 
\begin{eqnarray}
&&\stackrel{03}{(+i)}\stackrel{12}{(+)}|\stackrel{56}{(+)}\stackrel{78}{(+)}
||\stackrel{9 \;10}{(+)}\stackrel{11\;12}{(-)}\stackrel{13\;14}{(-)} |\psi \rangle =
\nonumber\\
&&(\gamma^0 -\gamma^3)(\gamma^1 +i \gamma^2)| (\gamma^5 + i\gamma^6)(\gamma^7 +i \gamma^8)||
(\gamma^9 +i\gamma^{10})(\gamma^{11} -i \gamma^{12})(\gamma^{13}-i\gamma^{14})|\psi \rangle.
\nonumber\\
\label{start}
\end{eqnarray}

The signs "$|$" and "$||$" are to point out the  $SO(1,3)$ (up to $|$), $SO(1,7)$ (up to $||$)
and $SO(6)$ (after $||$) substructure of the starting basic vector of the left handed multiplet of
$SO(1,13)$, which has $2^{14/2-1}= 64 $ vectors. Here $|\psi\rangle$ is any vector, which is not 
transformed 
to zero and therefore we shall not write down $|\psi \rangle$ any longer.
One easily finds that the eigen values of the chosen 
Cartan subalgebra elements of $S^{ab}$ and $\tilde{S}^{ab}$ (Eq.(\ref{cartan})) 
are $(+i/2, 1/2, 1/2,1/2,1/2,-1/2,-1/2)$ 
and $(+i/2, 1/2, 1/2,1/2,1/2,-1/2,-1/2)$,  
respectively. This state has with respect to the operators $S^{ab}$ the following properties:  
With respect to the group $SO(1,3)$  is a right handed spinor  ($\Gamma^{(1,3)} =1$) 
 with spin up ($S^{12} =1/2$), it is weak chargeless (it is an $SU(2)$  
singlet - $\tau^{13} = 0$) and it carries a colour charge (it is the member 
of  the $SU(3)$ triplet with ($\tau^{33} =1/2, \tau^{38} = 1/(2 \sqrt{3})$),
it has $\tau^{21} = 1/2$ and $\tau^{41}= 1/6$ and correspondingly the two hyper charges 
equal to $Y=2/3$ and $Y'= -1/3$, respectively. We further find
according to Eq.(\ref{handedness}) that $\Gamma^{(4)} =1$ (the handedness of the group $SO(4)$, whose 
subgroups are $SU(2)$ and $U(1)$), 
$\Gamma^{(1,7)}= 1$ and $ \Gamma^{(6)} = -1$. The starting vector (Eq.(\ref{start})) can be recognized 
in terms of the Standard model subgroups as the right handed weak chargeless $u$-quark carrying 
one of the three colours.

To obtain all the basic vectors of one Weyl spinor, one only has to apply on the starting basic vector of Eq.(\ref{start})
the generators $S^{ab}$. All the quarks and the leptons of one family of the Standard model appear in this multiplet 
(together with the corresponding anti quarks and anti leptons). 
We present in Table I all the quarks of one particular colour (the right handed weak 
chargeless $u_R,d_R$ and left handed weak charged $u_L, d_L$, with the colour $(1/2,1/(2\sqrt{3}))$ in 
the  Standard model notation). They all are  members of one $SO(1,7)$ multiplet.

\begin{center}
\begin{tabular}{|r|c||c||c|c||c|c|c||c|c|c||r|r|}
\hline
i&$$&$|^a\psi_i>$&$\Gamma^{(1,3)}$&$ S^{12}$&$\Gamma^{(4)}$&
$\tau^{13}$&$\tau^{21}$&$\tau^{33}$&$\tau^{38}$&$\tau^{41}$&$Y$&$Y'$\\
\hline\hline
&& ${\rm Octet},\;\Gamma^{(1,7)} =1,\;\Gamma^{(6)} = -1,$&&&&&&&&&& \\
&& ${\rm of \; quarks}$&&&&&&&&&&\\
\hline\hline
1&$u_{R}^{c1}$&$\stackrel{03}{(+i)}\stackrel{12}{(+)}|\stackrel{56}{(+)}\stackrel{78}{(+)}
||\stackrel{9 \;10}{(+)}\stackrel{11\;12}{(-)}\stackrel{13\;14}{(-)}$
&1&1/2&1&0&1/2&1/2&$1/(2\sqrt{3})$&1/6&2/3&-1/3\\
\hline 
2&$u_{R}^{c1}$&$\stackrel{03}{[-i]}\stackrel{12}{[-]}|\stackrel{56}{(+)}\stackrel{78}{(+)}
||\stackrel{9 \;10}{(+)}\stackrel{11\;12}{(-)}\stackrel{13\;14}{(-)}$
&1&-1/2&1&0&1/2&1/2&$1/(2\sqrt{3})$&1/6&2/3&-1/3\\
\hline
3&$d_{R}^{c1}$&$\stackrel{03}{(+i)}\stackrel{12}{(+)}|\stackrel{56}{[-]}\stackrel{78}{[-]}
||\stackrel{9 \;10}{(+)}\stackrel{11\;12}{(-)}\stackrel{13\;14}{(-)}$
&1&1/2&1&0&-1/2&1/2&$1/(2\sqrt{3})$&1/6&-1/3&2/3\\
\hline 
4&$d_{R}^{c1}$&$\stackrel{03}{[-i]}\stackrel{12}{[-]}|\stackrel{56}{[-]}\stackrel{78}{[-]}
||\stackrel{9 \;10}{(+)}\stackrel{11\;12}{(-)}\stackrel{13\;14}{(-)}$
&1&-1/2&1&0&-1/2&1/2&$1/(2\sqrt{3})$&1/6&-1/3&2/3\\
\hline
5&$d_{L}^{c1}$&$\stackrel{03}{[-i]}\stackrel{12}{(+)}|\stackrel{56}{[-]}\stackrel{78}{(+)}
||\stackrel{9 \;10}{(+)}\stackrel{11\;12}{(-)}\stackrel{13\;14}{(-)}$
&-1&1/2&-1&-1/2&0&1/2&$1/(2\sqrt{3})$&1/6&1/6&1/6\\
\hline
6&$d_{L}^{c1}$&$\stackrel{03}{(+i)}\stackrel{12}{[-]}|\stackrel{56}{[-]}\stackrel{78}{(+)}
||\stackrel{9 \;10}{(+)}\stackrel{11\;12}{(-)}\stackrel{13\;14}{(-)}$
&-1&-1/2&-1&-1/2&0&1/2&$1/(2\sqrt{3})$&1/6&1/6&1/6\\
\hline
7&$u_{L}^{c1}$&$\stackrel{03}{[-i]}\stackrel{12}{(+)}|\stackrel{56}{(+)}\stackrel{78}{[-]}
||\stackrel{9 \;10}{(+)}\stackrel{11\;12}{(-)}\stackrel{13\;14}{(-)}$
&-1&1/2&-1&1/2&0&1/2&$1/(2\sqrt{3})$&1/6&1/6&1/6\\
\hline
8&$u_{L}^{c1}$&$\stackrel{03}{(+i)}\stackrel{12}{[-]}|\stackrel{56}{(+)}\stackrel{78}{[-]}
||\stackrel{9 \;10}{(+)}\stackrel{11\;12}{(-)}\stackrel{13\;14}{(-)}$
&-1&-1/2&-1&1/2&0&1/2&$1/(2\sqrt{3})$&1/6&1/6&1/6\\
\hline\hline
\end{tabular}
\end{center}
Table I. The 8-plet of quarks - the members of $SO(1,7)$ subgroup, belonging to one Weyl left 
handed ($\Gamma^{(1,13)} = -1 = \Gamma^{(1,7)} \times \Gamma^{(6)}$) spinor representation of 
$SO(1,13)$. 
It contains the left handed weak charged quarks and the right handed weak chargeless quarks of a particular 
colour ($(1/2,1/(2\sqrt{3}))$). Here  $\Gamma^{(1,3)}$ defines the handedness in $(1+3)$ space, 
$ S^{12}$ defines the ordinary spin (which can also be read directly from the basic vector), 
$\tau^{13}$ defines the weak charge, $\tau^{21}$ defines the $U(1)$ charge, $\tau^{33}$ and 
$\tau^{38}$ define the colour charge and $\tau^{41}$ another $U(1)$ charge, which together with the
first one defines $Y$ and $Y'$. 
The reader can find the whole Weyl representation in the ref.\cite{Portoroz03}.

In Table II we present the leptons of one family of the Standard model. All the leptons belong 
to the same multiplet with respect to the group $SO(1,7)$. 
They are colour chargeless and differ accordingly from the quarks in Table I in the 
second $U(1)$ charge and in the colour charge.  The quarks and the leptons are 
equivalent with respect to the group $SO(1,7)$.  

\begin{center}
\begin{tabular}{|r|c||c||c|c||c|c|c||c|c|c||r|r|}
\hline
i&$$&$|^a\psi_i>$&$\Gamma^{(1,3)}$&$ S^{12}$&$\Gamma^{(4)}$&
$\tau^{13}$&$\tau^{21}$&$\tau^{33}$&$\tau^{38}$&$\tau^{41}$&$Y$&$Y'$\\
\hline\hline
&& ${\rm Octet},\;\Gamma^{(1,7)} =1,\;\Gamma^{(6)} = -1,$&&&&&&&&&& \\
&& ${\rm of \; leptons}$&&&&&&&&&&\\
\hline\hline
1&$\nu_{R}$&$\stackrel{03}{(+i)}\stackrel{12}{(+)}|\stackrel{56}{(+)}\stackrel{78}{(+)}
||\stackrel{9 \;10}{(+)}\stackrel{11\;12}{[+]}\stackrel{13\;14}{[+]}$
&1&1/2&1&0&1/2&0&$0$&-1/2&0&-1\\
\hline 
2&$\nu_{R}$&$\stackrel{03}{[-i]}\stackrel{12}{[-]}|\stackrel{56}{(+)}\stackrel{78}{(+)}
||\stackrel{9 \;10}{(+)}\stackrel{11\;12}{[+]}\stackrel{13\;14}{[+]}$
&1&-1/2&1&0&1/2&0&$0$&-1/2&0&-1\\
\hline
3&$e_{R}$&$\stackrel{03}{(+i)}\stackrel{12}{(+)}|\stackrel{56}{[-]}\stackrel{78}{[-]}
||\stackrel{9 \;10}{(+)}\stackrel{11\;12}{[+]}\stackrel{13\;14}{[+]}$
&1&1/2&1&0&-1/2&0&$0$&-1/2&-1&0\\
\hline 
4&$e_{R}$&$\stackrel{03}{[-i]}\stackrel{12}{[-]}|\stackrel{56}{[-]}\stackrel{78}{[-]}
||\stackrel{9 \;10}{(+)}\stackrel{11\;12}{[+]}\stackrel{13\;14}{[+]}$
&1&-1/2&1&0&-1/2&0&$0$&-1/2&-1&0\\
\hline
5&$e_{L}$&$\stackrel{03}{[-i]}\stackrel{12}{(+)}|\stackrel{56}{[-]}\stackrel{78}{(+)}
||\stackrel{9 \;10}{(+)}\stackrel{11\;12}{[+]}\stackrel{13\;14}{[+]}$
&-1&1/2&-1&-1/2&0&0&$0$&-1/2&-1/2&-1/2\\
\hline
6&$e_{L}$&$\stackrel{03}{(+i)}\stackrel{12}{[-]}|\stackrel{56}{[-]}\stackrel{78}{(+)}
||\stackrel{9 \;10}{(+)}\stackrel{11\;12}{[+]}\stackrel{13\;14}{[+]}$
&-1&-1/2&-1&-1/2&0&0&$0$&-1/2&-1/2&-1/2\\
\hline
7&$\nu_{L}$&$\stackrel{03}{[-i]}\stackrel{12}{(+)}|\stackrel{56}{(+)}\stackrel{78}{[-]}
||\stackrel{9 \;10}{(+)}\stackrel{11\;12}{[+]}\stackrel{13\;14}{[+]}$
&-1&1/2&-1&1/2&0&0&$0$&-1/2&-1/2&-1/2\\
\hline
8&$\nu_{L}$&$\stackrel{03}{(+i)}\stackrel{12}{[-]}|\stackrel{56}{(+)}\stackrel{78}{[-]}
||\stackrel{9 \;10}{(+)}\stackrel{11\;12}{[+]}\stackrel{13\;14}{[+]}$
&-1&-1/2&-1&1/2&0&0&$0$&-1/2&-1/2&-1/2\\
\hline\hline
\end{tabular}
\end{center}
Table II. The 8-plet of leptons - the members of $SO(1,7)$ subgroup, belonging to one Weyl left 
handed ($\Gamma^{(1,13)} = -1 = \Gamma^{(1,7)} \times \Gamma^{(6)}$) spinor representation of 
$SO(1,13)$, is presented. 
It contains  the left handed weak charged leptons and  the right handed weak chargeless leptons, all 
 colour chargeless. The two $8$-plets in Table I and II are equivalent with respect to the groups  
 $SO(1,7)$. They only differ in properties with respect to the group $SU(3)$ and $U(1)$ and consequently 
 in $Y$ and $Y'$.


\subsection{Appearance of families}
\label{techniquefamilies}

While the generators of the Lorentz group $S^{ab}$, with  a pair of $(ab)$, 
which does not belong
to the Cartan subalgebra (Eq.(\ref{cartan})), transform one vector of one Weyl 
representation into another vector of the same Weyl representation, 
transform the generators $\tilde{S}^{ab}$ (again if the pair $(ab)$ does not belong to the Cartan set) 
a member of one family into the same member of another family, leaving all the other quantum numbers 
(determined by $S^{ab}$)
unchanged\cite{norma92,norma93,norma95,norma01,holgernorma00,Portoroz03}. 
This is happening since  the application of  $\gamma^a$ (from the left)  changes the operator 
$\stackrel{ab}{(+)}$ (or the operator $\stackrel{ab}{(+i)}$)
into the operator $\stackrel{ab}{[-]}$ (or the operator $\stackrel{ab}{[-i]}$, respectively), 
while the operator
$\tilde{\gamma}^a$ (which is understood, up to a factor $\pm i$, as the application of 
$\gamma^a$ from the right hand side) 
changes $\stackrel{ab}{(+)}$ (or $\stackrel{ab}{(+i)}$)  into $\stackrel{ab}{[+]}$
(or into $\stackrel{ab}{[+i]}$, respectively), without changing the ''eigen values'' of the Cartan 
subalgebra set of the operators $S^{ab}$. Bellow, as an example, the 
application of $\tilde{S}^{01}$ on the 
state of Eq.({\ref{start}}) (up to a constant) is presented:
\begin{eqnarray}
\stackrel{03}{(+i)} \stackrel{12}{(+)}| \stackrel{56}{(+)} \stackrel{78}{(+)}||
\stackrel{9 10}{(+)} \stackrel{11 12}{(-)} \stackrel{13 14}{(-)} \nonumber\\
\stackrel{03}{[+i]} \stackrel{12}{[+]}| \stackrel{56}{(+)} \stackrel{78}{(+)}|| 
\stackrel{9 10}{(+)} \stackrel{11 12}{(-)} \stackrel{13 14}{(-)}.
\label{twofamilies}
\end{eqnarray}
One can easily see that both vectors of (\ref{twofamilies})  describe a right handed $u$-quark of 
the same colour. They are equivalent with respect to the operators $S^{ab}$. They only differ
in properties, determined by the operators $\tilde{S}^{ab}$  and accordingly also with respect to the 
Cartan subalgebra set  ($\tilde{S}^{03}, \tilde{S}^{12},\cdots,\tilde{S}^{13\;14}$). The first row vector 
has the following ''eigen values'' of this Cartan subalgebra set  
$(i/2,1/2,1/2,1/2,1/2,-1/2,-1/2)$, while the corresponding ''eigen values'' of the 
vector in the second row 
are $(-i/2,-1/2,1/2,1/2,1/2,-1/2,-1/2)$. Therefore, the operators $\tilde{S}^{ab}$ 
can be used to generate 
families of quarks and leptons of  the Standard model.

We present here some useful relations, concerning families 
\begin{eqnarray}
\tilde{S}^{ac}\stackrel{ab}{(k)}\stackrel{cd}{(l)}& = &\frac{i}{2} \eta^{aa} \eta^{cc} 
\stackrel{ab}{[k]}\stackrel{cd}{[l]}, \quad\;\;
\tilde{S}^{ad}\stackrel{ab}{(k)}\stackrel{cd}{(l)}  =  \frac{1}{2} l \eta^{aa}  
\stackrel{ab}{[k]}\stackrel{cd}{[l]},  \nonumber\\
\tilde{S}^{bc}\stackrel{ab}{(k)}\stackrel{cd}{(l)} &=& \frac{1}{2} k \eta^{cc}  
\stackrel{ab}{[k]}\stackrel{cd}{[l]},\quad \quad\;\;
\tilde{S}^{bd}\stackrel{ab}{(k)}\stackrel{cd}{(l)}  =  -\frac{i}{2} k l  
\stackrel{ab}{[k]}\stackrel{cd}{[l]},
\nonumber\\
\tilde{S}^{ac}\stackrel{ab}{[k]}\stackrel{cd}{[l]}& = & -\frac{i}{2}  
\stackrel{ab}{(k)}\stackrel{cd}{(l)}, \quad \quad \;\;\;\;\;
\tilde{S}^{ad}\stackrel{ab}{[k]}\stackrel{cd}{[l]}  =   \frac{1}{2} l \eta^{cc}
\stackrel{ab}{(k)}\stackrel{cd}{(l)}, \nonumber\\
\tilde{S}^{bc}\stackrel{ab}{[k]}\stackrel{cd}{[l]}& = & \frac{1}{2} k \eta^{aa}  
\stackrel{ab}{(k)}\stackrel{cd}{(l)}, \quad \quad\;\;
\tilde{S}^{bd}\stackrel{ab}{[k]}\stackrel{cd}{[l]}  =   \frac{i}{2} kl \eta^{aa} \eta^{cc} 
\stackrel{ab}{(k)}\stackrel{cd}{(l)}, 
\nonumber\\
\tilde{S}^{ac}\stackrel{ab}{(k)}\stackrel{cd}{[l]}& = & -\frac{i}{2} \eta^{aa}  
\stackrel{ab}{[k]}\stackrel{cd}{(l)}, \quad \;\;\;\;
\tilde{S}^{ad}\stackrel{ab}{(k)}\stackrel{cd}{[l]}  =    \frac{1}{2} l \eta^{aa} \eta^{cc}   
\stackrel{ab}{[k]}\stackrel{cd}{(l)}, \nonumber\\
\tilde{S}^{bc}\stackrel{ab}{(k)}\stackrel{cd}{[l]}& = & -\frac{1}{2} k   
\stackrel{ab}{[k]}\stackrel{cd}{(l)}, \quad\quad \;\;\;\;
\tilde{S}^{bd}\stackrel{ab}{(k)}\stackrel{cd}{[l]}  =   -\frac{i}{2} k l \eta^{cc}  
\stackrel{ab}{[k]}\stackrel{cd}{(l)},
\nonumber\\
\tilde{S}^{ac}\stackrel{ab}{[k]}\stackrel{cd}{(l)}& = &\frac{i}{2} \eta^{cc}  
\stackrel{ab}{(k)}\stackrel{cd}{[k]}, \quad \quad \;\;\;\;
\tilde{S}^{ad}\stackrel{ab}{[k]}\stackrel{cd}{(l)}  =  \frac{1}{2} l  
\stackrel{ab}{(k)}\stackrel{cd}{[k]},\nonumber\\
\tilde{S}^{bc}\stackrel{ab}{[k]}\stackrel{cd}{(l)}& = &- \frac{1}{2} k \eta^{aa} \eta^{cc}  
\stackrel{ab}{(k)}\stackrel{cd}{[k]}, \;\;
\tilde{S}^{bd}\stackrel{ab}{[k]}\stackrel{cd}{(l)}  =  \frac{i}{2} k l \eta^{aa}  
\stackrel{ab}{(k)}\stackrel{cd}{[k]}.
\label{tildesac}
\end{eqnarray}
%


\section{Mass matrices in the Approach unifying spins and charges - terms within each family}
\label{Yukawawithin}

We are now prepared to look at the terms, which manifest as the Yukawa couplings in our approach  
unifying spins and charges. Let us at first neglect the terms $\tilde{S}^{ab}\tilde{\omega}_{abc}$ 
(Eqs.(\ref{lagrange},\ref{yukawa})) to which $\tilde{S}^{ab}$ contribute and see how does the ordinary  
Poincar\' e gauging gravity contribute to
the Yukawa couplings. It contributes to the matrix elements  within families only.

Let us look at the Lagrange density (Eq.(\ref{yukawa})) for one Weyl spinor of a
particular handedness - say the left handed one ($\Gamma^{(1+7)}=-1$) - and of all possible 
$SU(3)\times U(1)$ charges. 
The Lagrange density (Eq.(\ref{yukawa})) manifests  couplings of a spinor with the colour, 
the weak and the hyper charges 
fields and it also manifests the mass term   
\begin{eqnarray}
{\mathcal L} &=& \bar{\psi} \gamma^{m} (p_{m}- \sum_{A,i} \; g^{A} 
\tau^{Ai} A^{Ai}_{m}) \; \psi + \nonumber\\
& & + \psi^+ \gamma^0 \gamma^s \;(p_s - \sum_{A,i} \; g^{A} 
\tau^{Ai} A^{Ai}_{s}  ) \psi
+ {\rm  terms  \; with}\;  \tilde{S}^{ab}\tilde{\omega}_{abc} + {\rm the \; rest}.
\label{yukawa1}
\end{eqnarray}
We use the notation $f^{\alpha}{}_{a} p_{\alpha} = p_a$.  Since for simplicity we assume that in the 
''physical space'' there is no gravity, it follows: $f^{\mu}{}_{m} = 
\delta^{\mu}{}_{m}$. We also (just) assume that the term of the type $\tilde{\tau}^{Ai} \tilde{A}^{Ai}_{m}) $ 
is negligible at "physical energies".  
We recognize that the terms with $\gamma^0 \gamma^7$ or $\gamma^0 \gamma^8$ transform a right 
handed weak chargeless spinors with the 
spin $1/2$ (like it is the  $u_R$ quark from the first row in Table I or the $e_R$ electron from the 
third row in Table II) into a left handed weak charged spinors with the spin $1/2$ (in our example into  
the  $u_L$ quark from the seventh's row in Table I or the  $e_L$ electron from the fifth's row in Table II).

We can rearrange the first term in the Lagrangean of Eq.(\ref{yukawa1}), with $m \in \{0,1,2,3\},$  
to  manifest the Standard model structure
\begin{eqnarray}
{\mathcal L}_{f} = \bar{\psi} \gamma^{m} \{p_{m} &-& \frac{g}{2} (\tau^{+} W^{+}_{m} +
\tau^{-} W^{-}_{m})  + \frac{g^2}{\sqrt{g^2 +g'^2}} Q' Z_{m} + \nonumber\\ 
&+& \frac{g g'}{\sqrt{g^2 +g'^2}} Q A_{m}  +\nonumber\\
 &+&  \sum_{i} g^3 \tau^{3i} A^{3i}_{m}  + \; g^{Y'}  A^{Y'}_{m}\}\; \psi, 
\label{lagrangesmf}
\end{eqnarray}
with 
\begin{eqnarray}
Q  &=&  \tau^{13} + Y = S^{56} + \tau^{41}, \nonumber\\
Q' &=& \tau^{13} - (\frac{g'}{g})^2 Y = \frac{1}{2} 
(1- (\frac{g'}{g})^2) S^{56} - \frac{1}{2}(1+ (\frac{g'}{g})^2) S^{78} - (\frac{g'}{g})^2 \tau^{41}.
\label{Q}
\end{eqnarray}
We assume that (due to an appropriate  break of symmetries) the term $A^{Y'}_{m} $ is non observable at 
''physical'' energies (not yet).  

We rearrange    the mass term of the Lagrange density of Eq.(\ref{yukawa1})     
in a similar way as the ''dynamical'' part of the Lagrange density in the ''physical space'' 
(the first term of this equation),  
so that the fields $A^{Ai}_{s}, s\in \{5,6,7,8\}$, 
instead of $\omega_{t t' s} $, appear.
The charge $Q$ is conserved, as seen in Eq.(\ref{Q}),  if we assume that no terms with 
either $\gamma^5$ or $\gamma^6$ or $\tau^{3i}$
 contribute to the mass term.

Since  all the operators in Eq.(\ref{yukawa1}) are to be applied on right handed spinors,  
which are weak chargeless objects (as seen from Table I and II), the part with $\sum_{i} \; 
\tau^{1i} A^{1i}_{s} $ contributes zero to mass matrices and we shall leave it out. 
We also expect,  that at the ''observable'' energies the contribution of 
the components $p_s$ of momenta, with $s=5,6,\cdots, $   
are negligible. Accordingly we neglect also this term.  

What then stays in the Lagrange density for the mass terms of spinors if the terms with 
$\tilde{S}^{ab}$ are not yet taken into account, is as follows\footnote{We shall from now on simplify  
the notation 
from $g^{A} A^{Ai}_{a}$ to $A^{Ai}_{a}$.}
\begin{eqnarray}
- {\mathcal L}_{Y} &=& - \psi^+ \gamma^0 \sum_{s=7,8} \gamma^s \; ( 
Y A^{Y}_{s} +  
Y' A^{Y'}_{s} ) \psi \nonumber\\
&+& {\rm  terms  \; with}\;  \tilde{S}^{ab}\tilde{\omega}_{abc}.
\label{yukawa3}
\end{eqnarray}
These terms distinguish among the spinors: they are different for quarks than for leptons, 
as well as different 
for the 
$u$ quarks  than for the $d$ quarks and different for  the electrons than for the neutrinos, 
since according to Table I and II,  
different spinors carry different values of the two hyper charges. 
The first two terms in Eq.(\ref{yukawa3}) 
contribute to mass matrices within one family only. 
The expression for the mass terms (the Yukawa couplings in the Standard model language) 
within one family (Eq.(\ref{yukawa3})) can be further rewritten, 
if introducing the following superposition of operators 
\begin{eqnarray}
(\gamma^7 \pm i \gamma^8 ) = 2 \stackrel{78}{(\pm)}. 
\label{pm}
\end{eqnarray}
It then follows
\begin{eqnarray}
- {\mathcal L}_{Y} &=& - \psi^+ \gamma^0 \; \sum_{y=Y,Y'}\; 
\{ \stackrel{78}{(+)} \; y A^{y}_{+}\; + \; \stackrel{78}{(-)} \; y A^{y}_{-} \}\psi \nonumber\\
&+& {\rm  terms  \; with}\;  \tilde{S}^{ab}\tilde{\omega}_{abc},
\label{yukawa4}
\end{eqnarray}
with $A^{y}_{\pm} = - (A^{y}_{7} \mp i A^{y}_{8})$ and $y=Y,Y'$. 
According to Eq.(\ref{graphbinoms}), 
saying that 
\begin{eqnarray}
\stackrel{78}{(+)} \stackrel{78}{(+)} &=& 0, \quad \quad \quad \stackrel{78}{(-)} \stackrel{78}{(+)} \;
= \; - \stackrel{78}{[-]}, \nonumber\\
\stackrel{78}{(+)} \stackrel{78}{[-]} &=&  \stackrel{78}{(+)}, \quad \quad \;\stackrel{78}{(-)} 
\stackrel{78}{[-]} \; = \; 0,
\label{graphbinomssel}
\end{eqnarray}
we conclude, after reading also Table I and Table II,  that the term with the 
fields $A^{y}_{+}$, $y=Y, Y'$,  
contributes only to the masses of the $d-$quarks and the electrons, 
while $A^{y}_{-}$, $y=Y, Y'$, contributes 
only to  the masses of the $u-$quarks and the neutrinos.

\section{Mass matrices in the Approach unifying spins and charges - terms within and among families}
\label{Yukawafamilies}

We have seen in Subsect.\ref{techniquefamilies} that while the operators $S^{ab}$ transform the 
members of one Weyl representation among themselves, the operators $\tilde{S}^{ab}$ transform one 
member of a family into the same member of another family, changing nothing but the family index.
Each spinor basic vector has accordingly two indices: one index tells to
which family a spinor belongs, another index tells which member of a particular family a spinor 
represents. 

There are two types of terms in the Lagrange density of Eq.(\ref{lagrange}), which contribute to the 
mass matrices. We have studied in the previous Sect.\ref{Yukawawithin} only the terms, determined by  
the generators of the Lorentz group ($S^{ab}$) and the corresponding gauge fields. After making a few 
assumptions we ended up with quite a 
simple expression for the contribution to {\em the masses of quarks and leptons within one family}. 

The assumption, that there are two kinds of gauge fields connected with two kinds of the generators 
of the Lorentz transformations, is new\cite{norma93,holgernorma00,norma01,pikanormaproceedings2} and
requires accordingly additional cautions, when using it. On the other hand, the idea of the 
existence of {\em two kinds of Clifford algebra objects 
leads to the concept of families} and  is therefore too exciting 
not to be used  to try to describe families of quarks and leptons 
and to see what can the approach say about families of quarks and leptons. 

As already said, the two kinds of generators are in our approach accompanied by the 
two kinds of gauge fields, 
 gauging $S^{ab}$ and $\tilde{S}^{ab}$, respectively. 
We shall assume  that breaks of the symmetries of 
the Poincar\' e group in $d=1+13$ influence this additional spin connection field as well. 
Since we do not know, how these breaks could occur for any of these two kinds of degrees of freedom, 
 the calculations which will follow can be understood only as an attempt to  study these degrees 
of freedom, that is families of quarks and leptons, within the proposed approach and
and not yet as a severe prediction of properties of families of this approach.

On the other side, many a property of families, presented in this paper, not connected 
with the break of symmetries, can follow from any concept of 
construction of families,  if  operators for generating families  
commute with the generators of spins and charges and if the generators are accompanied by 
gauge fields. What our proposal might offer  
in addition to the general concept of families is the prediction of the number of families and 
possible relations among  matrix elements of mass matrices due to connected effects followed by 
breaks of symmetries.

Our approach suggests an even number of families. Namely: The number of all the orthogonal 
basic states is in our approach for a particular $d$ equal to $2^d$. Since we start  
with one Weyl spinor and  neither $S^{ab}$ nor $\tilde{S}^{ab}$ change 
the Clifford  oddness or evenness of  basic states, we stay with $2^{d-2}$ basic states. Each Weyl 
representation has $2^{d/2-1}$ members. For any  $d$ there are accordingly $2^{d/2-1}$ families. 
After the assumption that some spontaneous breaks of symmetries lead to massless spinors in $d=(1+7)$-
dimensional space, in  which spinors carry the $SU(3)$ and $U(1)$ charges (the two $U(1)$'s, one from  
$SO(6)$ and another from $SO(1,7)$, being connected), there are accordingly only
$2^{8/2-1} = 8$ families at most, seen at low energies. The Yukawa-like terms further break symmetry, 
leading to $Q= S^{56} + \tau^{41}$ and $SO(1,3), SU(3)$ as  conserved quantities and to massive spinors  
in the ''physical'' space with $d=(1+3)$. 

Since the application of $\tilde{S}^{ab}$ generates the equivalent representations with respect to the 
Lorentz group ($\{S^{ab},\tilde{S}^{cd}\}_{-} 
= 0$ and accordingly $ \tilde{S}^{ab} \stackrel{ab}{(k)} = \tilde{S}^{ab} \stackrel{ab}{[-k]}$), it 
follows that if we know properties with respect to $\tilde{S}^{ab}$ for one of basic states of  
a Weyl spinor, 
we know  them for the whole Weyl spinor - {\em up to the influence of the break of symmetries} and 
up to the fact that the contribution to the Yukawa couplings of the two kinds of generators are 
in our approach related.

To demonstrate properties of families we shall make use of the first 
state in Table I and Table II. The two tables differ only in the $SU(3)$ and $ U(1)$ 
charges (both kinds originate in $SO(6)$) and these two 
charges do not concern the $SO(1,7)$ part. 
Accordingly we shall tell only families, connected with $SO(1,7)$. 
\begin{eqnarray}
I.\;\;& & \stackrel{03}{(+i)} \stackrel{12}{(+)} |\stackrel{56}{(+)} \stackrel{78}{(+)}||...\quad 
V.\; \stackrel{03}{[+i]} \stackrel{12}{(+)} |\stackrel{56}{(+)} \stackrel{78}{[+]}||... \nonumber\\
II.\;& &\stackrel{03}{[+i]} \stackrel{12}{[+]} |\stackrel{56}{(+)} \stackrel{78}{(+)}||... \quad
VI. \stackrel{03}{(+i)} \stackrel{12}{[+]} |\stackrel{56}{(+)} \stackrel{78}{[+]}||... \nonumber\\
III.& &\stackrel{03}{(+i)} \stackrel{12}{(+)} |\stackrel{56}{[+]} \stackrel{78}{[+]}||...\quad
VII. \stackrel{03}{(+i)} \stackrel{12}{[+]} |\stackrel{56}{[+]} \stackrel{78}{(+)}||... \nonumber\\
IV.& &\stackrel{03}{[+i]} \stackrel{12}{[+]} |\stackrel{56}{[+]} \stackrel{78}{[+]}||...\quad 
VIII.\stackrel{03}{[+i]} \stackrel{12}{(+)} |\stackrel{56}{[+]} \stackrel{78}{(+)}||.... .
\label{eightfamilies}
\end{eqnarray}
The diagonal terms, to which $S^{st}\omega_{sts'}$ contribute, depend on a state of the 
Weyl representation, 
as we have commented in Sect.\ref{Yukawawithin}, and distinguish between quarks and 
leptons, as well as 
between the $u$ and  the $d$ quarks and between the electrons and  the neutrinos. 
The diagonal and the 
off  diagonal elements, to which $\tilde{S}^{ab} \tilde{\omega}_{abs}$ contribute,  
 distinguish  between the $u$ and the $d$ quarks and between the electrons and the neutrinos, due to 
 the factor $\stackrel{78}{(\pm)}$ (determined by the ordinary $\gamma^a$ operators), but do not 
 distinguish between the quarks and the leptons. However,  the way of breaking 
 symmetries  might influence the commutation relations among 
 both kinds of generators. 
 (We could, in some rough estimation, take these effects into account for instance just by putting an 
 additional index to gauge 
 fields $\tilde{\omega}_{abc}$.  We shall discuss such possibilities in the paper which will 
 present the numerical results.)
 The diagonalization of the $u$ ($\nu$) mass matrix 
 leads accordingly  to a different transformation
matrix than the diagonalization of $d$ ($e$) mass matrix. The mixing matrix 
for quarks is correspondingly 
not the unit matrix (as expected, if it should agree with the experimental data) but might not differ  
from the mixing matrix for leptons.

Let us now write down the whole expression for the Yukawa couplings, with 
$\tilde{S}^{ab} \tilde{\omega}_{ab \alpha}$ 
included 
\begin{eqnarray}
- {\mathcal L}_{Y} &=& \psi^{\dagger} \gamma^0 \gamma^s p_{0 s} \psi\;  = \psi^{\dagger} \; \gamma^0 \{
 \stackrel{78}{(+)} \; p_{0+}\; + \; \stackrel{78}{(-)} \;p_{0-} \}\psi,
\label{yukawatildeyukawa}
\end{eqnarray}
with 
\begin{eqnarray}
p_{0\pm} = (p_{7} \mp i \; p_{8}) - \frac{1}{2} S^{ab} \omega_{ab\pm}  - 
\frac{1}{2} \tilde{S}^{ab} \tilde{\omega}_{ab\pm},  \nonumber\\
\omega_{ab\pm} = \omega_{ab7} \mp i \; \omega_{ab8}, \quad 
\tilde{\omega}_{ab\pm} = \tilde{\omega}_{ab 7} \mp i \; \tilde{\omega}_{ab8}. 
\label{yukawatildeyukawadet}
\end{eqnarray}

We shall rewrite diagonal matrix elements, to which  $\tilde{S}^{ab}$ contribute, in a similar way as 
we did in the previous sections for the contribution of $S^{ab}$. 
We therefore introduce the appropriate superposition of the operators 
$\tilde{S}^{ab}$ 
\begin{eqnarray}
\tilde{N}^{\pm}_{3}: &=& \frac{1}{2} ( \tilde{{\mathcal S}}^{12} \pm i
\tilde{{\mathcal S}}^{03} ),  \nonumber\\
\tilde{\tau}^{13} : &=& \frac{1}{2} ( \tilde{{\mathcal S}}^{56} - \tilde{{\mathcal S}}^{78} ),  \nonumber\\
\tilde{Y} &=& \tilde{\tau}^{41} + \tilde \tau^{21}, \quad  
\tilde{Y'} = \tilde{\tau}^{41} - \tilde{\tau}^{21}, \nonumber\\
\tilde{\tau}^{21}: &=& \frac{1}{2} ( \tilde{{\mathcal S}}^{56} + \tilde{{\mathcal S}}^{78} ), \quad
\tilde{\tau}^{41}: = -\frac{1}{3} ( \tilde{{\mathcal S}}^{9 \;10} + \tilde{{\mathcal S}}^{11\; 12} 
+  \tilde{{\mathcal S}}^{13\; 14} ).
\label{tildetau}
\end{eqnarray}
We allow also terms with $\tilde{S}^{mn}$,$ m,n=0,1,2,3,$ which in diagonal matrix 
elements of a mass matrix appear as $\tilde{N}^{\pm}_{3}$. 
Taking into account that
\begin{eqnarray}
-\frac{1}{2} S^{st} \omega_{st\pm} &=& Y A^{Y}_{\pm} + Y' A^{Y'}_{\pm} + \tau^{13} A^{13}_{\pm}, \nonumber\\
-\frac{1}{2} \tilde{S}^{st} \tilde{\omega}_{st\pm} &=& \tilde{Y} \tilde{A}^{\tilde{Y}}_{\pm} + 
\tilde{Y'} \tilde{A}^{\tilde{Y'}}_{\pm} + \tilde{\tau}^{13} \tilde{A}^{13}_{\pm}, \nonumber\\
-\frac{1}{2} \tilde{S}^{mn}\tilde{\omega}_{mn\pm} &=& \tilde{N}^{+3} \tilde{A}^{+3}_{\pm} +
\tilde{N}^{-3} \tilde{A}^{-3}_{\pm},
\label{tildetaufields}
\end{eqnarray}
with the pairs $(m,n) =(0,3),(1,2)$; $(s,t) = (5,6),(7,8),$ belonging to the Cartan sub algebra and 
$\Omega_{\pm} = \Omega_7 \mp i \Omega_8$, where $\Omega_7, \Omega_8$ 
stay for any of the above fields, we find
\begin{eqnarray}
A^{13}_{\pm} &=& - (\omega_{56\pm} - \omega_{78 \pm}), \nonumber\\
A^{Y}_{\pm}  &=& -\frac{1}{2} (A^{41}_{\pm} + (\omega_{56\pm} + \omega_{78\pm})),\nonumber\\
A^{Y'}_{\pm} &=& -\frac{1}{2} (A^{41}_{\pm} - (\omega_{56\pm} + \omega_{78\pm})), \nonumber\\
\tilde{A}^{13}_{\pm} &=& - (\tilde{\omega}_{56\pm} - \tilde{\omega}_{78 \pm}), \nonumber\\
\tilde{A}^{\tilde{Y}}_{\pm}  &=& -\frac{1}{2} (\tilde{A}^{41}_{\pm} + (\tilde{\omega}_{56\pm}
+ \tilde{\omega}_{78\pm})),\nonumber\\
\tilde{A}^{\tilde{Y'}}_{\pm} &=& -\frac{1}{2} (\tilde{A}^{41}_{\pm} - (\tilde{\omega}_{56\pm} + 
\tilde{\omega}_{78\pm})), \nonumber\\
\tilde{A}^{\tilde{N}^{+}_{3}}_{\pm} &=& - (\tilde{\omega}_{12 \pm} - i \; \tilde{\omega}_{03\pm}),\nonumber\\
\tilde{A}^{\tilde{N}^{-}_{3}}_{\pm} &=& - (\tilde{\omega}_{12 \pm} + i \; \tilde{\omega}_{03\pm}),
\label{tildetaufields1}
\end{eqnarray}
where the fields $A^{y}_{\pm},\; y= 13,41,Y,Y',$ and $\tilde{A}^{\tilde{y}}_{\pm},\; \tilde{y}
= \tilde{N}^{+}_{3},
\tilde{N}^{-}_{3}, 13, 41,
\tilde{Y},\tilde{Y'},$ are uniquely expressible with the 
corresponding spin connection fields. Let us repeat that $\omega_{ab c} = 
f^{\alpha}{}_{c} \;\omega_{ab \alpha}$ and $\tilde{\omega}_{ab c} = f^{\alpha}{}_{c} \;
\tilde{\omega}_{ab \alpha}$.

The operators, which contribute to non diagonal terms in mass matrices, are superpositions of 
$\tilde{S}^{ab}$ and can be written in terms of nilpotents 
\begin{eqnarray}
\stackrel{ab}{\tilde{(k)}}\stackrel{cd}{\tilde{(l)}},  
\label{lowertilde}
\end{eqnarray}
with indices $(ab)$ and $(cd)$ which belong to the Cartan sub algebra indices (Eq.(\ref{cartan})).
We may write accordingly
\begin{eqnarray}
  \sum_{(a,b) } -\frac{1}{2} \stackrel{78}{(\pm)}\tilde{S}^{ab} \tilde{\omega}_{ab\pm} =
- \sum_{(ac),(bd), \;  k,l}\stackrel{78}{(\pm)}\stackrel{ac}{\tilde{(k)}}\stackrel{bd}{\tilde{(l)}} 
\; \tilde{A}^{kl}_{\pm} ((ac),(bd)),  
\label{lowertildeL}
\end{eqnarray}
where the pair $(a,b)$ in the first sum runs over all the  indices, which do not characterise  
the Cartan sub algebra, with $ a,b = 0,\dots, 8$,  while the two pairs $(ac)$ and $(bd)$ 
denote only the Cartan sub algebra pairs
 (for $SO(1,7)$ we only have the pairs $(03), (12)$; $(03), (56)$ ;$(03), (78)$;
$(12),(56)$; $(12), (78)$; $(56),(78)$ ); $k$ and $l$ run over four 
possible values so that $k=\pm i$, if $(ac) = (03)$ 
and $k=\pm 1$ in all other cases, while $l=\pm 1$.
The fields  $\tilde{A}^{kl}_{\pm} ((ac),(bd))$ can then 
be expressed by $\tilde{\omega}_{ab \pm}$ as follows 
\begin{eqnarray}
\tilde{A}^{++}_{\pm} ((ab),(cd)) &=& -\frac{i}{2} (\tilde{\omega}_{ac\pm} -\frac{i}{r} \tilde{\omega}_{bc\pm} 
-i \tilde{\omega}_{ad\pm} -\frac{1}{r} \tilde{\omega}_{bd\pm} ), \nonumber\\
\tilde{A}^{--}_{\pm} ((ab),(cd)) &=& -\frac{i}{2} (\tilde{\omega}_{ac\pm} +\frac{i}{r} \tilde{\omega}_{bc\pm} 
+i \tilde{\omega}_{ad\pm} -\frac{1}{r} \tilde{\omega}_{bd\pm} ),\nonumber\\
\tilde{A}^{-+}_{\pm} ((ab),(cd)) &=& -\frac{i}{2} (\tilde{\omega}_{ac\pm} + \frac{i}{r} \tilde{\omega}_{bc\pm} 
-i  \tilde{\omega}_{ad\pm} +\frac{1}{r} \tilde{\omega}_{bd\pm} ), \nonumber\\
\tilde{A}^{+-}_{\pm} ((ab),(cd)) &=& -\frac{i}{2} (\tilde{\omega}_{ac\pm} - \frac{i}{r} \tilde{\omega}_{bc\pm} 
+i  \tilde{\omega}_{ad\pm} +\frac{1}{r} \tilde{\omega}_{bd\pm} ),
\label{Awithomega}
\end{eqnarray}
with $r=i$, if $(ab) = (03)$ and $r=1$ otherwise.
 We simplify the index $kl$ in the exponent 
of fields $\tilde{A}^{kl}{}_{\pm} ((ac),(bd))$ to $\pm $, omitting $i$. 

Any break of symmetries in the $\tilde{S}^{ab}$ sector would cause relations among the 
corresponding $\tilde{\omega}_ {ab\pm}$. Namely, if $-\frac{1}{2}\tilde{S}^{ab} 
\tilde{\omega}_{ ab\pm} = \tilde{\tau}^{Ai} \tilde{A}^{Ai}_{\pm}$ is not just 
a unitary transformation of basic states, but means due to a break of symmetries that, 
let us say, a particular  $\tilde{A}^{A'i}=0$, then this can only happen, if 
$\tilde{\omega}_{ab \pm}$ are related. 

The Lagrange density, representing the mass matrices of fermions 
(the Yukawa couplings in the Standard model) 
(Eq.(\ref{yukawatildeyukawa})), can  be rewritten as follows
\begin{eqnarray}
{\mathcal L}_{Y} = \psi^+ \gamma^0 \;  
\{ & &\stackrel{78}{(+)} ( \sum_{y=Y,Y'}\; y A^{y}_{+} + 
\sum_{\tilde{y}=\tilde{N}^{+}_{3},\tilde{N}^{-}_{3},\tilde{\tau}^{13},\tilde{Y},\tilde{Y'}} 
\tilde{y} \tilde{A}^{\tilde{y}}_{+}\;)\; + \nonumber\\
  & & \stackrel{78}{(-)} ( \sum_{y=Y,Y'}\;y  A^{y}_{-} +  
\sum_{\tilde{y}= \tilde{N}^{+}_{3},\tilde{N}^{-}_{3},\tilde{\tau}^{13},\tilde{Y},\tilde{Y'}} 
\tilde{y} \tilde{A}^{\tilde{y}}_{-}\;) + \nonumber\\
 & & \stackrel{78}{(+)} \sum_{\{(ac)(bd) \},k,l} \; \stackrel{ac}{\tilde{(k)}} \stackrel{bd}{\tilde{(l)}}
\tilde{{A}}^{kl}_{+}((ac),(bd)) \;\;+  \nonumber\\
 & & \stackrel{78}{(-)} \sum_{\{(ac)(bd) \},k,l} \; \stackrel{ac}{\tilde{(k)}}\stackrel{bd}{\tilde{(l)}}
\tilde{{A}}^{kl}_{-}((ac),(bd))\}\psi,
\label{yukawa4tilde}
\end{eqnarray}
with  pairs $((ac),(bd))$, which run over all the members of the Cartan sub algebra, while  $k=\pm i,$ if 
$ (ac) =(03)$, otherwise $k=\pm 1$ and $l= \pm 1$. 
The terms $\tilde{{A}}^{kl}_{\pm}((ac),(bd))$ are expressible in terms of $\tilde{\omega}_{ab \pm}$ 
as presented in Eq.(\ref{Awithomega}), while any break of symmetries relates $\tilde{\omega}_{ab \pm}$ 
in a very particular way. 
We omitted the term with $\tau^{13}$, as well as the terms 
$p_{\pm}$, since, as already explained in the previous section, the first one when being applied 
on the right handed spinors contributes zero, while for the second ones we assume that at 
low energies their contribution is negligible. 

The mass matrix, which follows from these Lagrange density and  depends strongly on all possible breaks 
of symmetries, is in general not  Hermitean.

Let us now repeat the assumptions we have made up to now. They are either the starting assumptions 
of our approach unifying spins and charges, or  we made them  to be able to connect the starting 
Lagrange density  at low energies with the observable phenomena.

a.i. We use the approach, unifying spins and charges, which assumes, that in $d=1+13$ massless 
spinors carry  two types of 
spins: the ordinary (in $d=1 + 13$) one, which we describe by $S^{ab} = 
\frac{1}{4}(\gamma^a \gamma^b - \gamma^b \gamma^a)$ and the additional one, 
described by $\tilde{S}^{ab} =
\frac{1}{4}(\tilde{\gamma}^a \tilde{\gamma}^b - \tilde{\gamma}^b \tilde{\gamma}^a)$. 
The two types of the 
Clifford algebra objects anti commute ($\{\gamma^a, \tilde{\gamma}^b \}_+ =0$). 
Spinors carry no charges in $d=1+13$. 
The operators $S^{ab}$ determine (after an appropriate break of symmetries) at low energies the ordinary 
spin in $d= 1+3$ 
and all the known charges, while $\tilde{S}^{ab}$ generate families of spinors. Accordingly spinors 
interact with only the gravitational fields, the gauge fields of the Poincar\' e group ($p_{\alpha}$, 
$S^{ab}$), and the gauge fields of the operators $\tilde{S}^{ab}$ 
($p_{0a} = f^{\alpha}{}_{a} - \frac{1}{2} (S^{cd} \omega_{cda} + \tilde{S}^{cd} \tilde{\omega}_{cda})).$

a.ii. The break of symmetries of $SO(1,13)$ into $SO(1,7)\times SU(3)\times U(1)$ 
occurs in a way that only massless spinors 
in $d=1+7$ with the charge $ SU(3)$ and $ U(1)$ survive, with the one $U(1)$ from $SO(1,7)$ 
and the next $U(1)$ 
from $SO(6)$ aligned, while $S^{56}$ does not contribute to the Yukawa-like terms, so that 
$Q= \tau^{41} + S^{56}$ is conserved in $d=1+3$.

a.iii. The break of symmetries  influences both: the Poincar\' e symmetry and the symmetry 
described by $\tilde{S}^{ab}$, it might be that to some extend in a similar way. 
The study of both kinds of breaking 
symmetries stays as an open problem.

a.iv. The terms which include $p_{s}, s = 5,..,14,$ do not 
contribute at low energies.


%
\section{An example of mass matrices for  four families}
\label{example}

Let us make, for simplicity, two further assumptions besides the four (a.i-a.iv.) ones, 
presented at the end of Sect.\ref{Yukawafamilies}:

b.i. There are no terms, which would in Eq.(\ref{eightfamilies}) transform  
$\stackrel{\tilde{56}}{(+)}$ into $\stackrel{\tilde{56}}{[+]}$. 
This assumption (which could also be understood as a break of symmetry, which 
requires that terms of the type $\tilde{S}^{5a}\tilde{\omega}_{5ab}$ and 
$\tilde{S}^{6a}\tilde{\omega}_{6ab}$ are negligible and might be  a part of 
requirement a.iii.)  
leaves us with only four families 
of quarks and leptons. (This assumption might be justified with a break of symmetry 
in the $\tilde{S}^{ab}$ sector from $SO(1,7)$ to $SO(1,5) \times U(1)$, with  all 
the contributions of the terms $\tilde{S}^{5a}\tilde{\omega}_{5ab}$ and 
$\tilde{S}^{6a}\tilde{\omega}_{6ab}$ equal to zero.)
 
b.ii. The rough estimation will be done on ''a tree level''.

Since we do not know either how does the break of symmetries occur or how 
does the break influence the 
strength of the fields $\omega_{abc}$ and $\tilde{\omega}_{abc}$,
we can not really say, to which extend are the above assumptions justified. For 
none of them  we have a justification. Also the nonperturbative effects could be very 
strong and the tree level might not mean a lot. 
But yet  a simplified version can help 
us to understand to what conclusions might the proposed approach lead with respect to 
families of quarks and leptons and their properties.

Our approach (which predicts an even number of families) suggests that under the assumptions a. and b.  
there are the following four families of quarks and leptons
\begin{eqnarray}
I.\;& & \stackrel{03}{(+i)} \stackrel{12}{(+)} |\stackrel{56}{(+)} \stackrel{78}{(+)}||...\nonumber\\ 
II.\;& &\stackrel{03}{[+i]} \stackrel{12}{[+]} |\stackrel{56}{(+)} \stackrel{78}{(+)}||... \nonumber\\
III.& & \stackrel{03}{[+i]} \stackrel{12}{(+)} |\stackrel{56}{(+)} \stackrel{78}{[+]}||... \nonumber\\
IV. & & \stackrel{03}{(+i)} \stackrel{12}{[+]} |\stackrel{56}{(+)} \stackrel{78}{[+]}||... .
\label{fourfamilies}
\end{eqnarray}
We see from Table I (and II) that due to the properties of the nilpotents $\stackrel{78}{(\pm)}$ 
(Eq.\ref{graphbinoms}),
to the $u$ quark (and to the $\nu$ lepton) mass matrix only the operator $\stackrel{78}{(-)}$ 
(accompanied by the fields $A_-, \tilde{A}_-$) 
contributes, while to the $d$ quark (and to the $e$ lepton) mass matrix only $\stackrel{78}{(+)}$ 
(accompanied by the fields $A_+, \tilde{A}_+$) contributes. 
This means  that {\em the off diagonal matrix elements of the Yukawa couplings are  different 
for $u$-quarks ($\nu$) and for $d$-quarks (e)}, although still related, while the quarks have 
the same off diagonal matrix elements as the corresponding leptons (unless some breaks 
of symmetries do not destroys this symmetry).  
Assuming that after the appropriate breaks of symmetries the fields 
contributing to the Yukawa couplings obtain some nonzero expectation values (which are in general 
related in a very particular way) and 
integrating the Lagrange density $L_Y$ over the coordinates and the internal (spin) degrees of freedom, 
we end up   
with the mass matrices for four families of quarks and leptons (Eq.(\ref{fourfamilies})), whose  
structure is presented in Table III. 
\begin{center}
\begin{tabular}{|r||c|c|c|c|}
\hline
$$&$ I_{R} $&$ II_{R} $&$ III_{R} $&$ IV_{R}$\\
\hline\hline
&&&& \\
$I_{L}$   & $ A^I_{\mp} $ & $ \tilde{A}^{++}_{\mp} ((03),(12)) $ & $ \pm \tilde{A}^{++}_{\mp} ((03),(78))$ &
$ \mp  \tilde{A}^{++}_{\mp} ((12),(78))$ \\
&&&& \\
\hline 
&&&& \\
$II_{L}$  & $ \tilde{A}^{--}_{\mp} ((03),(12)) $ & $ A^{II}_{\mp} $ & $ \pm \tilde{A}^{-+}_{\mp} ((12),(78))$ &
$ \mp  \tilde{A}^{-+}_{\mp} ((03),(78))$ \\
&&&& \\
\hline 
&&&& \\
$III_{L}$ & $ \pm \tilde{A}^{--}_{\mp} ((03),(78)) $ & $\mp \tilde{A}^{+-}_{\mp} ((12),(78)) $ & $ A^{III}_{\mp}$ &
$  \tilde{A}^{-+}_{\mp} ((03),(12))$ \\
&&&& \\
\hline 
&&&& \\
$IV_{L}$  & $\pm \tilde{A}^{--}_{\mp} ((12),(78)) $ & $\mp \tilde{A}^{+-}_{\mp} ((03),(78)) $ & 
$ \tilde{A}^{+-}_{\mp} ((03),(12))$ & $ A^{IV}_{\mp} $ \\
&&&& \\
\hline\hline
\end{tabular}
\end{center}
Table III. The mass matrices for four families of quarks and leptons in the approach unifying spins 
and charges, obtained under the assumptions a.i.- a.iv. and b.i.- b.ii..
The values $ A^{I'}_{-}, I'= I,II,III,IV,$ and $ \tilde{A}^{lm}_{-} ((ac),(bd)); l,m = \pm,$  determine 
matrix elements for the $u$ quarks and the neutrinos, the values $ A^{I'}_{+},  I'= I,II,III,IV,$ and $
\tilde{A}^{lm}_{+} ((ac),(bd)); l,m =\pm,$  determine  the matrix elements for the $d$ quarks and 
the electrons. 
Diagonal matrix elements are different for quarks than for  leptons and distinguish also between  the $u$ 
and the $d$ quarks and between the $\nu$ and the $e$ leptons (Eqs.\ref{yukawa4tilde}, 
\ref{tildetaufields1}). 
They also differ from family to family. Non diagonal matrix elements distinguish  among families and 
among $(u, \nu)$ and $(d,e)$. 
The presented  matrix should  be understood as a very preliminary  estimate of the mass matrices 
of quarks and leptons. 

The explicit forms of the diagonal matrix elements for the above choice of assumptions in terms of 
$\omega_{abc}$, 
$\tilde{\omega}_{abc}$  and $\tilde{A}^{41}_{\pm}$ is  as follows
\begin{eqnarray}
A^{I}_{u} &=& \frac{2}{3} A^{Y}_{-} - \frac{1}{3} A^{Y'}_{-}  + \tilde{\omega}^{I}_{-},
\nonumber\\
A^{I}_{\nu} &=& - A^{Y'}_{-}  + \tilde{\omega}^{I}_{-}, 
 \nonumber\\
A^{I}_{d} &=& -\frac{1}{3} A^{Y}_{+} + \frac{2}{3} A^{Y'}_{+}  + \tilde{\omega}^{I}_{+}, 
 \nonumber\\
A^{I}_{e} &=& - A^{Y}_{+} +  \tilde{\omega}^{I}_{+},
 \nonumber\\
A^{II}_{u} &= & A^{I}_{u} + (i \tilde{\omega}_{03-} + \tilde{\omega}_{12 -}), \quad
A^{II}_{\nu} =  A^{I}_{\nu} + (i \tilde{\omega}_{03-} + \tilde{\omega}_{12 -}), \nonumber\\
A^{II}_{d} &= & A^{I}_{d} + (i \tilde{\omega}_{03+} + \tilde{\omega}_{12 +}), \quad
A^{II}_{e} =  A^{I}_{e} + (i \tilde{\omega}_{03+} + \tilde{\omega}_{12 +}), \nonumber\\
A^{III}_{u} &= & A^{I}_{u} + (i \tilde{\omega}_{03-} + \tilde{\omega}_{78 -}), \quad
A^{III}_{\nu} =  A^{I}_{\nu} + (i \tilde{\omega}_{03-} + \tilde{\omega}_{78 -}), \nonumber\\
A^{III}_{d} &= & A^{I}_{d} + (i \tilde{\omega}_{03+} + \tilde{\omega}_{78 +}), \quad
A^{III}_{e} =  A^{I}_{e} + (i \tilde{\omega}_{03+} + \tilde{\omega}_{78 +}), \nonumber\\
A^{IV}_{u} &= & A^{I}_{u} + ( \tilde{\omega}_{12-} + \tilde{\omega}_{78 -}), \quad
A^{IV}_{\nu} =  A^{I}_{\nu} + ( \tilde{\omega}_{12-} + \tilde{\omega}_{78 -}), \nonumber\\
A^{IV}_{d} &= & A^{I}_{d} + ( \tilde{\omega}_{12+} + \tilde{\omega}_{78 +}), \quad
A^{IV}_{e} = A^{I}_{e} + ( \tilde{\omega}_{12+} + \tilde{\omega}_{78 +}),
\label{diagmfour}
\end{eqnarray}
with $-
\tilde{\omega}^{I}{}_{\pm} = \frac{1}{2} (i \tilde{\omega}_{03\pm} + 
\tilde{\omega}_{12\pm} +\tilde{\omega}_{56\pm} + \tilde{\omega}_{78\pm}
+ \frac{1}{3} \tilde{A}^{41}_{\pm})$.
The explicit forms of non diagonal matrix elements are written in Eq.(\ref{Awithomega}). 
As allready stated, the break of symmetries, which is not taken into account in Table III,  
would strongly relate ''vacuum expactation values'' of $\tilde{\omega}_{ab \pm}$.

To evaluate briefly the structure of mass matrices we make one further assumption:

b.iii. Let the mass matrices be real and symmetric (while all the $\omega_{abc}$ and 
$\tilde{\omega}_{abc}$ are assumed to be real). 

We then obtain for the $u$ quarks (and neutrinos) the  mass matrices as presented in Table IV.

\begin{center}
\begin{tabular}{|r||c|c|c|c|}
\hline
$u$&$ I_{R} $&$ II_{R} $&$ III_{R} $&$ IV_{R}$\\
\hline\hline
&&&& \\
$I_{L}$   & $ A^I_{u}  $ & $ \tilde{A}^{++}_{u} ((03),(12))= $ & $  \tilde{A}^{++}_{u} ((03),(78)) =$  &
$ -  \tilde{A}^{++}_{u} ((12),(78)) = $ \\
$$&$$ &$ \frac{1}{2}(\tilde{\omega}_{327} +\tilde{\omega}_{018})$&$ \frac{1}{2}(\tilde{\omega}_{387} +\tilde{\omega}_{078}) $&$
 \frac{1}{2}(\tilde{\omega}_{277} +\tilde{\omega}_{187})$ \\
&&&& \\
\hline
&&&&\\
$II_{L}$  & $ \tilde{A}^{--}_{u} ((03),(12))= $ & $ A^{II}_{u}= $ & $  \tilde{A}^{-+}_{u} ((12),(78)) = $ &
$ -  \tilde{A}^{-+}_{u} ((03),(78)) = $ \\
$$&$ \frac{1}{2}(\tilde{\omega}_{327} +\tilde{\omega}_{018})$&$A^{I}_{u} +  (\tilde{\omega}_{127} - \tilde{\omega}_{038})$
&$ -\frac{1}{2}(\tilde{\omega}_{277} -\tilde{\omega}_{187}) $&$  \frac{1}{2}(\tilde{\omega}_{387} - \tilde{\omega}_{078})$ \\
\hline 
&&&& \\
$III_{L}$ & $  \tilde{A}^{--}_{u} ((03),(78)) =$ & $- \tilde{A}^{+-}_{u} ((12),(78))= $ & $ A^{III}_{u}=$ &
$   \tilde{A}^{-+}_{u} ((03),(12)) =$ \\
$$&$  \frac{1}{2}(\tilde{\omega}_{387} +\tilde{\omega}_{078})$&$ -\frac{1}{2}(\tilde{\omega}_{277} -\tilde{\omega}_{187})$
&$A^{I}_{u} +  (\tilde{\omega}_{787} - \tilde{\omega}_{038})$&$ -\frac{1}{2}(\tilde{\omega}_{327} -\tilde{\omega}_{018}) $\\
&&&& \\
\hline 
&&&& \\
$IV_{L}$  & $ \tilde{A}^{--}_{u} ((12),(78)) =$ & $- \tilde{A}^{+-}_{u} ((03),(78)) = $ & 
$ \tilde{A}^{+-}_{u} ((03),(12))$ & $ A^{IV}_{u} =$ \\
$$&$ \frac{1}{2}(\tilde{\omega}_{277} +\tilde{\omega}_{187})$&$ \frac{1}{2}(\tilde{\omega}_{387} -\tilde{\omega}_{078})$
&$ -\frac{1}{2}(\tilde{\omega}_{327} -\tilde{\omega}_{018}) $&$A^{I}_{u} +  (\tilde{\omega}_{127} + \tilde{\omega}_{787})$\\
&&&& \\
\hline\hline
\end{tabular}
\end{center}
Table IV. The mass matrix of four families of the $u$-quarks (and neutrinos) obtained within the approach 
unifying spins and charges and under the assumptions a.i.-a.iv., b.i.-b.iii..
Neutrinos and $u$-quarks distinguish in   $ A^I_{u} \ne  A^I_{\nu} $.  
The break of symmetries, not yet taken into account, would relate $\tilde{\omega}_{ab 7,8}$ and would 
accordingly reduce the number of free parameters. 

The corresponding mass matrix for $d$-quarks  (and electrons) is presented in Table V.

\begin{center}
\begin{tabular}{|r||c|c|c|c|}
\hline
$d-$&$ I_{R} $&$ II_{R} $&$ III_{R} $&$ IV_{R}$\\
\hline\hline
&&&& \\
$I_{L}$   & $ A^I_{d}  $ & $ \tilde{A}^{++}_{d} ((03),(12))= $ & $  - \tilde{A}^{++}_{d} ((03),(78)) =$  &
$   \tilde{A}^{++}_{d} ((12),(78)) = $ \\
$$&$$ &$ \frac{1}{2}(\tilde{\omega}_{327} - \tilde{\omega}_{018})$&$ -\frac{1}{2}(\tilde{\omega}_{387} - 
\tilde{\omega}_{078}) $&$  -\frac{1}{2}(\tilde{\omega}_{277} +\tilde{\omega}_{187})$ \\
&&&& \\
\hline
&&&&\\
$II_{L}$  & $ \tilde{A}^{--}_{d} ((03),(12))= $ & $ A^{II}_{d}= $ & $  -\tilde{A}^{-+}_{d} ((12),(78)) = $ &
$  \tilde{A}^{-+}_{d} ((03),(78)) = $ \\
$$&$ \frac{1}{2}(\tilde{\omega}_{327} -\tilde{\omega}_{018})$&$A^{I}_{d} +  (\tilde{\omega}_{127} + \tilde{\omega}_{038})$
&$ \frac{1}{2}(\tilde{\omega}_{277} -\tilde{\omega}_{187}) $&$  -\frac{1}{2}(\tilde{\omega}_{387} +\tilde{\omega}_{078})$ \\
\hline 
&&&& \\
$III_{L}$ & $  - \tilde{A}^{--}_{d} ((03),(78)) =$ & $ \tilde{A}^{+-}_{d} ((12),(78))= $ & $ A^{III}_{d}=$ &
$   \tilde{A}^{-+}_{d} ((03),(12)) =$ \\
$$&$ -\frac{1}{2}(\tilde{\omega}_{387} -\tilde{\omega}_{078})$&$ \frac{1}{2}(\tilde{\omega}_{277} -\tilde{\omega}_{187})$
&$A^{I}_{d} +  (\tilde{\omega}_{787} + \tilde{\omega}_{038})$&$ -\frac{1}{2}(\tilde{\omega}_{018} +\tilde{\omega}_{327}) $\\
&&&& \\
\hline 
&&&& \\
$IV_{L}$  & $ -\tilde{A}^{--}_{d} ((12),(78)) =$ & $ \tilde{A}^{+-}_{d} ((03),(78)) = $ & 
$ \tilde{A}^{+-}_{d} ((03),(12))$ & $ A^{IV}_{d} $ \\
$$&$ -\frac{1}{2}(\tilde{\omega}_{277} +\tilde{\omega}_{187})$&$ -\frac{1}{2}(\tilde{\omega}_{387} +\tilde{\omega}_{078})$
&$ -\frac{1}{2}(\tilde{\omega}_{018} +\tilde{\omega}_{327}) $&$A^{I}_{d} +  (\tilde{\omega}_{127} + \tilde{\omega}_{787})$\\
&&&& \\
\hline\hline
\end{tabular}
\end{center}
Table V. The mass matrix of four families of the $d$-quarks and electrons. The
quarks and the leptons distinguish in this approximation in $ A^I_{d} \ne  A^I_{e}$. 
Other comments are the same as in Table IV.

The relation between the mass matrix of  $u$-quarks and the mass matrix of neutrinos is, 
under the assumptions and simplifications made during  deriving both tables, as follows: 
All the off-diagonal elements are for the neutrinos the same as for the $u$-quarks, while
the diagonal matrix elements depend on   
the eigen values of $Y$ and $Y'$. Accordingly, 
both $A^{I}_{\alpha},
\alpha =u,\nu,$ can be understood as independent parameters, expressible in terms of 
$A^{Y}_{-}, A^{Y'}_{-}, $ and $
\tilde{\omega}^{I}{}_{-}$.

 Similarly, 
 the relation between the mass matrix of $d$-quarks and the mass matrix for  electrons is, 
 under the same assumptions and simplification as used for finding the expressions for the mass 
 matrices for the $u-$quarks and the neutrinos, as follows:  
All the off-diagonal elements are the same for both - the $d$-quarks and the electrons, while
the diagonal matrix elements distinguish in 
the eigen values of $Y$ and $Y'$.
Again, both $A^{I}_{\beta},
\beta =d,e,$ can be understood as independent parameters, expressible in terms of 
$A^{Y}_{+}, A^{Y'}_{+}, $ and $
\tilde{\omega}^{I}{}_{+}$.

The requirement about  reality and symmetry of 
mass matrices, relates $A^{Y}_{+} = A^{Y}_{-}, A^{Y'}_{+} = A^{Y'}_{-}, 
\tilde{\omega}^{I}{}_{+} = \frac{1}{2} \tilde{\omega}_{03 8} + \tilde{\omega},
\tilde{\omega}^{I}{}_{-} = -\frac{1}{2} \tilde{\omega}_{03 8} + \tilde{\omega}$, 
where $\tilde{\omega} = \tilde{\omega}_{12 7} + 
\tilde{\omega}_{56 7} + \tilde{\omega}_{78 7}
+ \frac{1}{3} \tilde{A}^{41}$ and $\tilde{A}^{41}$ is the real part of either $\tilde{A}^{41}_{+} 
$ or $\tilde{A}^{41}_{-} $. The same  assumption  relates also off diagonal 
elements for $u$-quarks and $d$-quarks (or neutrinos and electrons), 
as seen from both tables, so that there are  $13$ free parameters, which determine $4\times 4(4+1)/2$ mass 
matrix elements, and from these mass matrices $4 \times 4$ masses of quarks and leptons and 
$2 (4(4+1)/2-1) $ 
elements of the two mixing matrices should follow.

Further break of symmetries would further  
relate the  $\tilde{\omega}_{ab \pm}$ fields, reducing strongly the number of free parameters on 
Table IV and  Table V.
A very peculiar boundary conditions could - when breaking symmetries - even cause  differences  
in off diagonal matrix elements  of  
quarks and leptons. Also could the nonperturbative effects  beyond the ''tree level'' be responsible  
for the differences observed in the measured  properties of quarks and leptons or for  
what in many references are trying  
to achieve with additional Higgs fields.  We did not take into account any Majorana fermions. 

Most of the above assumptions were proposed  to be able to make a rough 
estimation of properties of  the  mass matrices, 
predicted by the approach unifying spins and charges.

We shall present the calculations with the parameters presented in Tables IV and V in the paper, 
which will follow this one.

\section{Concluding discussions}
\label{conclusions}

In this paper we discuss about a possible origin of the families of quarks and leptons and 
of their Yukawa couplings as proposed by the approach unifying spins and 
charges\cite{norma92,norma93,normasuper94,%
norma95,norma97,pikanormaproceedings1,holgernorma00,norma01,%
pikanormaproceedings2,Portoroz03}.

The approach assumes that a Weyl spinor of a chosen handedness carries in 
$d (=1+13)-$ dimensional space nothing but two kinds of spin degrees of freedom. One kind belongs to 
the Poincar\' e group in $d=1+13$, 
another one  generates families. 
The idea of generating families with the second kind of the Clifford algebra objects 
(which commute  with the generators of the Lorentz transformations for spinors) is new, as it is 
new also the idea that there are the generators of the Lorentz transformations (accompanied by the 
spin connection fields in $d > 4$) which are (together with $ \gamma^0$)  
responsible for the Yukawa couplings within a family, 
transforming a right handed weak chargeless quark or lepton into a left handed weak charged 
one.  
Spinors interact with only the gravitational fields, 
manifested by  vielbeins and  spin connections, the gauge fields of the momentum 
$p_{\alpha}$ and the two kinds of the generators of the Lorentz group $S^{ab}$ and $\tilde{S}^{ab}$, 
respectively.

To derive the mass matrices from the starting Lagrangean - that is to calculate the Yukawa couplings  
of the Standard model - no additional (Higgs) field is needed. In order to make a simple and 
transparent evaluation of properties of the mass matrices and consequently 
some estimations and rough predictions for the 
masses and mixing matrices for quarks 
and leptons, observed at ''physical'' energies, 
we made several  assumptions, approximations and simplifications, 
not necessary all of them are ''physical'' (and some of them 
should soon  be relaxed in further studies):

i.  The break of symmetries of the group $SO(1,13)$ into $SO(1,7)\times SU(3)\times U(1)$ occurs 
in a way that only 
massless spinors in $d=1+7$ with the charge $ SU(3)\times U(1)$ survive. And yet 
the two $U(1)$ charges, 
following from $SO(6)$ and $SO(1,7)$, respectively, are related. 
(Our work on the compactification 
of a massless spinor in $d=1+5$ into   $d=1+3$ and a finite disk gives us 
some hope that this assumption 
might be fulfilled\cite{holgernorma05}.) 
The requirement that the terms 
with $S^{5a}, S^{6a}$  do 
not contribute to the 
mass term at low energies, assures that the charge  
$Q= \tau^{41} + S^{56}$ is conserved.

ii. The break of symmetries influences the Poincar\' e symmetry and the symmetries described by 
$\tilde{S}^{ab}$.  But it is assumed that the gauge symmetries connected with $\tilde{S}^{ab}$ 
do not manifest  as  gauge fields (additional to the known charge gauge fields) in $d=1+3$.  
It is also assumed that there are no terms, which would in Eq.(\ref{eightfamilies}) transform  
$\stackrel{\tilde{56}}{(+)}$ into $\stackrel{\tilde{56}}{[+]}$ (which can be explained by the 
related break of symmetries in  $S^{ab}$ and $\tilde{S}^{ab}$ sektor). This assumption reduces the 
number of families by a factor $2$. It really means that the group $SO(1,7)$, whose generators are 
$\tilde{S}^{ab}$, is broken into $SO(1,5)\times U(1)$ in a way that terms 
$\tilde{S}^{5a}\tilde{\omega}_{5ab}$ and  $\tilde{S}^{6a}\tilde{\omega}_{6ab}$ bring no 
contribution to the mass matrices. Otherwise no additional break of symmetry was taken into account. 

We also assume that terms which 
include the components $p_s, s=5,..,14$  of the momentum $p^a$ do not 
contribute at low energies to the mass matrices. We leave for further studies to find out 
how do different ways of breaking symmetries of the Poincar\' e group in $d=1+13$ and the 
$SO(1,13)$ group of $\tilde{S}^{ab}$ influence the mass matrices.

iii. We  make estimations on a ''tree level''.  

iv. We assume the mass matrices to be real and symmetric.

Our starting Weyl spinor representation of a chosen handedness in $d(=1+13)-$dimensional 
space manifests, if analyzed in terms of the 
subgroups $SO(1,3), SU(3), SU(2)$ and two $U(1)'$s (the sum of the ranks of the subgroups is the 
same as the rank of the starting group) of the group $SO(1,13)$, the spin and all the charges 
of one family of quarks and leptons. It includes left handed weak charged quarks and leptons 
and right handed weak chargeless quarks and leptons in the same representation 
and does accordingly {\em answer one of 
the open questions of the Standard model: Why only the left handed fermions carry the 
weak charge while the right handed ones are weak chargeless}, how can it at all happen 
that handedness, which concerns only the spin (in $d=1+3$), is so strongly related to
a (weak) charge? 

We use our technique\cite{holgernorma02,technique03} to present spinor representations 
in a transparent way so that one easily sees how does a part of the covariant 
derivative of a spinor in $d=1+13$ manifest in $d=1+3$ as Yukawa couplings. 
We use the same technique to represent also families of spinors. Since the starting action in 
$d={1+13}$ manifests in $d=1+3$ the (even number) of families and the Yukawa couplings, 
it {\em offers a possible 
answer to the questions, why families of quarks and leptons and 
the corresponding Yukawa couplings manifest in nature}.

We found the off diagonal mass matrix elements  of the quarks and the  
leptons strongly related. 
We expect that these relations might very probably turn out to be  too strong  
(since there are  in our case the same off diagonal and diagonal matrix elements, 
which determine the orthogonal rotations  of the matrices for the $u-$quarks and neutrinos 
and the $d-$quarks and electron into the diagonal forms and 
consequently also the corresponding mixing matrices for quarks are the same as for leptons, 
while the experimental data 
shows quite a difference in the mixing matrices of  quarks and leptons). We suspect that some 
particular breaks of symmetries  (with the help of very peculiar 
boundary conditions added) might be responsible in our approach 
for the differences between quarks and leptons 
in   off diagonal 
and also those diagonal  matrix elements, which are generated    by the family generators.    
 Or might the reason for the difference  be in the Majorana like neutrinos (as suspected in 
 many references), which are not treated in these 
studies.     

If the symmetry of mass matrices as presented in this paper for four families 
breaks further, the relations among the parameters determining 
the mass matrices follow, reducing  the presented number of independent parameters 
 $\tilde{\omega}_{abcd}$. 
In particular,  an exact break of the symmetry of $SO(1,5)$ in the $\tilde{S}^{ab}$ 
sector into $SU(3)\times U(1)$ 
would manifest in decoupling of the fourth family from the first three (by 
relating $\tilde{\omega}_{abc}$ so that the corresponding matrix elements would be zero), 
while the break of 
$SO(1,5)$ into $SU(2)\times SU(2)\times U(1)$ would manifest in decoupling of the first 
two families from the second two families. 

We treat the quarks and the leptons in an equivalent way, with no Majorana neutrinos included.  

 Not all the above assumptions and simplifications are needed in order  
 to be able to estimate 
 mass matrices of quarks and leptons with not too much effort. And in addition, it might happen 
 that this too simplified estimates lead to unrealistic conclusions. (In particular, 
 the $CP$ violation can under 
 the assumption v. hardly  be possible.) One also can not expect, that ''a tree level'' estimate is 
 good enough  to evaluate properties of  quarks and leptons. Nonperturbative effects might 
 strongly influence the results and they  might even be  a very strong  reason for the difference 
 in properties 
 of the families of the $u$-quarks, the $d$-quarks and both kinds of leptons.   
 Corrections  bellow the  tree level might bring also the contributions, which  
 several references try to simulate by more than one Higgs 
 field.

For the four predicted families of quarks and leptons we present the explicit expressions 
for the mass matrices in the above mentioned approximations.

Since in our approach the generators of the Poincar\' e group and of those, 
generating families, commute, 
many a property of mass matrices, presented in this paper, would be true also for, let us say, 
models, in which 
the generators of the Poincar\' e group and those of generating families, commute. 
However, in our case  breaks of symmetries in the two sectors are related    
and these relations might be very important for the properties of quarks and leptons.   

On the other hand we should find the explanation why the additional gauge fields, connected 
with the $\tilde{S}^{ab}$ sector does not manifest in $d=1+3$. 

We shall present numerical estimates for the Yukawa couplings after 
relating our results with the known experimental data in the paper\cite{matjazdragannorma}, 
following this one, together with further 
discussions of the  properties of  families of quarks and leptons as following from  
the approach unifying spins and charges, in order to be able to see 
whether  this approach shows the right way 
beyond the Standard model of electroweak and colour interactions.

\section*{Acknowledgments} 
It is a pleasure to thank all the participants of the   workshops entitled 
"What comes beyond the Standard model", 
taking place  at Bled annually in  July, starting at 1998,  for many very fruitful discussions, 
in particular to H.B. Nielsen.

 \end{document}